\documentclass[10pt,twocolumn]{IEEEtran}
\usepackage{graphicx} 
\usepackage{epsfig} 
\usepackage{mathptmx} 
\usepackage{times} 
\usepackage{amsmath} 
\usepackage{amssymb}  
\usepackage{color}
\usepackage{authblk}

\usepackage[caption=false,font=footnotesize]{subfig}

\DeclareMathOperator*{\argmax}{argmax}

\newcommand{\textr}[1]{#1}

\newcommand{\ignore}[1]{}

\def\proof{\noindent{{\bf Proof: }}}
\def\endproof{\hspace*{\fill}~\QED\par\endtrivlist\unskip}
\newtheorem{theorem}{Theorem}
\newtheorem{lemma}{Lemma}

\newcommand{\E}[1]{{\mathbb{E}}\left[{#1}\right]}
\newcommand{\vc}[1]{{\mathbf{#1}}}

\newcommand{\Prob}[1]{{\mathbb{P}}\left({#1}\right)}
\newcommand{\test}{~ \begin{subarray}{c} {\cal I}_1^*=1 \\ \gtreqless \\ {\cal I}_1^*=0 \end{subarray} ~}

\newcommand{\Wv}{{\bf W}}
\newcommand{\Rv}{{\bf R}}

\begin{document}

\title{{Control of Wireless Networks with Secrecy}{\thanks{This material is based upon work supported by the National
Science Foundation under Grants CNS-0831919, CCF-0916664,
CAREER-1054738, and by Marie Curie International Research Staff
Exchange Scheme Fellowship PIRSES-GA-2010-269132 AGILENet within the
7th European Community Framework Programme.}}\thanks{Portions of
this work were presented at Asilomar Conference on Signals, Systems,
and Computers (Asilomar '10), Pacific Grove, CA.}}

\author{C.~Emre Koksal, Ozgur Ercetin, Yunus Sarikaya\vspace{-0.2in}%
\thanks{C.~E. Koksal (koksal@ece.osu.edu) is with the Department of Electrical and Computer Engineering at The Ohio State University, Columbus, OH.}  \thanks{O. Ercetin (email: oercetin@sabanciuniv.edu) and  Y.Sarikaya (email: sarikaya@su.sabanciuniv.edu) are with the Department of Electronics Engineering, Faculty of Engineering and Natural Sciences, Sabanci University, 34956 Istanbul, Turkey.}}

\maketitle

\begin{abstract}
We consider the problem of cross-layer resource allocation in
time-varying cellular wireless networks, and incorporate information
theoretic secrecy as a Quality of Service constraint. Specifically,
each node in the network injects two types of traffic, private and
open, at rates chosen in order to maximize a global utility
function, subject to network stability and secrecy constraints. The
secrecy constraint enforces an arbitrarily low mutual information
leakage from the source to every node in the network, except for the
sink node. We first obtain the achievable rate region for the
problem for single and multi-user systems assuming that the nodes
have full CSI of their neighbors. Then, we provide a joint flow
control, scheduling and private encoding scheme, which does not rely
on the knowledge of the prior distribution of the gain of any
channel. We prove that our scheme achieves a utility, arbitrarily
close to the maximum achievable utility. Numerical experiments are
performed to verify the analytical results, and to show the efficacy
of the dynamic control algorithm.
\end{abstract}
\vspace{-0.2in}
\section{Introduction}
\label{sec:intro}

In recent years, there have been a number of investigations on
wireless information theoretic secrecy. These studies have been
largely confined within the boundaries of the {\em physical layer}
in the wireless scenario and they have significantly enhanced our
understanding of the fundamental limits and principles governing the
design and analysis of secure wireless communication systems. For
example, \cite{fading,Barros:ISIT:06,Liang:TIT:061} have unveiled
the {\em opportunistic secrecy} principle which allows for
transforming the multi-path fading variations into a secrecy
advantage for the legitimate receiver, even when the eavesdropper is
enjoying a higher average signal-to-noise ratio (SNR). The
fundamental role of {\em feedback} in enhancing the secrecy capacity
of point-to-point wireless communication links was established
in~\cite{Lai:TIT:07,Ardestanizadeh:TIT:09,Gunduz:ISITA:08}. More
recent works have explored the use of~{\em multiple antennas} to
induce ambiguity at the eavesdropper under a variety of assumptions
on the available transmitter channel state information (CSI)
\cite{Khisti:TIT:07,Oggier:TIT:07,Shafiee:TIT:09,Liu:TIT:091}.
The multi-user aspect of the wireless environment was studied
in~\cite{Lai:TIT:061,Tekin:TIT:06,
Liang:TIT:08,Khisti:TIT:08,Bloch:TIT:08,Li:ITA:07,Simeone:CISS:09,
Parada:ISIT:05,Liu:ISIT:06,Oohama:TIT:06,Ekrem:TIT:09M,Yuksel:ITW:09,
Ali:TIFS:11,Li:TSP:11,Chen:TIFS:12}
revealing the potential gains that can be reaped from appropriately
constructed user cooperation policies. Finally, the design of
practical codes that approach the promised capacity limits was
investigated in~\cite{Liu:ITW:07,Bloch:ISIT:06}. One of the most
interesting outcomes of this body of work is the discovery of the
positive impacts on secure communications of some wireless
phenomena, e.g., interference, which are traditionally viewed as
impairments to be overcome.

Despite the significant progress in information theoretic secrecy,
most of the work has focused on physical layer techniques and on a
single link. The area of wireless information theoretic secrecy
remains in its infancy, especially as it relates to the design of
wireless networks and its impact on network control and protocol
development. Therefore, our understanding of the interplay between
the secrecy requirements and the critical functionalities of
wireless networks, such as {\em scheduling, routing, and congestion
control} remains very limited.

Scheduling in wireless networks is a prominent and challenging
problem which attracted significant interest from the networking
community.  The challenge arises from the fact that the capacity of
wireless channel is time varying due to multiple superimposed random
effects such as mobility and multipath fading. Optimal scheduling in
wireless networks has been extensively studied in the literature
under various assumptions \cite{tassiulas, shroff,
subramanian,urgaonkar, jaramillo, stolyar}.
Starting with
the seminal work of Tassiulas and Ephremides \cite{tassiulas} where
throughput optimality of backpressure algorithm is proven, policies
that opportunistically exploit the time varying nature of the
wireless channel to schedule users are shown to be at least as good
as static policies \cite{shroff}. In principle, these opportunistic
policies schedule the user with the favorable channel condition to
increase the overall performance of the system. However, without
imposing individual performance guarantees for each user in the
system, this type of scheduling results in unfair sharing of
resources and may lead to starvation of some users, for example,
those far away from the base station in a cellular network. Hence,
in order to address fairness issues, scheduling problem was
investigated jointly with the network utility maximization problem
\cite{kelly, Low,Kar},
and the stochastic network optimization framework \cite{Neely} was developed.

To that end, in this paper we address the basic wireless network
control problem in order to develop a cross-layer resource
allocation solution that will incorporate information privacy, {\em
measured by equivocation}, as a QoS metric. In particular, we
consider the single hop uplink setting, in which nodes collect
private and open information, store them in separate queues and
transmit them to the base station. At a given point in time, only
one node is scheduled to transmit and it may choose to transmit some
combination of open and private information. \textr{Our objective is
to achieve privacy of information from the other legitimate nodes
and we assume that there are no external malicious eavesdroppers in
the system. The motivation to study this notion of secrecy is the
following. In some scenarios (e.g., tactical, financial, medical),
privacy of communicated information between the nodes is necessary,
so that data intended to (or originated from) a node is not shared
by any other legitimate node.
}

First, we evaluate the region of achievable open and private data
rate pairs for a single node scenario with and without joint
encoding of open and private information. Then, we consider the
multi-node scenario, and introduce {\bf private opportunistic
scheduling}. We find the achievable private information rate regions
associated with private opportunistic scheduling and show that
\ignore{for both the uplink and the downlink scenarios,} it achieves
the maximum sum private information rate over all joint scheduling
and encoding strategies. \textr{ While private opportunistic
scheduler is based on the availability of full CSI on the uplink
channels, it does not rely on information on the instantaneous
cross-channel (i.e., the channel between different nodes) CSI. It
requires merely the long-term average rate of the cross-channel
rates. To achieve privacy with this level of CSI, private
opportunistic scheduler uses an encoding scheme that encodes private
information over many packets. Note that, in the seminal
paper~\cite{Knopp:ICC:95}, it was shown that opportunistic
scheduling (without secrecy) maximizes the sum rate. Our result can
be viewed as a generalization of this result to the case with
secrecy.}
Next, we model the problem as that of network utility maximization.
\textr{ We provide a dynamic joint flow control, scheduling and
private encoding scheme, which takes into account the
\emph{instantaneous} direct- and cross-channel state information but
not a priori channel state distribution. In dynamic cross-layer
control scheme private information is divided into a sequence of
messages where each message is encoded into an \emph{individual}
packet.} We prove that our scheme achieves a utility, arbitrarily
close to the maximum utility achievable in this setting. We
generalize dynamic cross-layer control scheme to a more general case
when instantaneous cross-channel states are not known perfectly.
Consequently, we define the notions of {\em privacy outage} and {\em
privacy goodput}. Finally, we numerically characterize the
performance of the dynamic control algorithm with respect to several
network parameters, and show that its performance is fairly close to
that of private opportunistic scheduler achievable with known
channel priors.

\vspace{-0.2in}
\section{Problem Model}
\label{sec:model}

\noindent
\begin{figure}
\centerline{\includegraphics[height=1.2in]{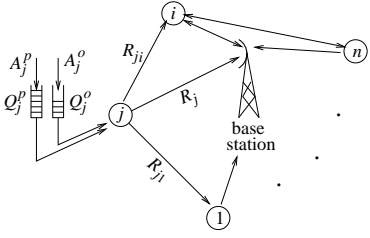}}
\caption{Uplink communication with private and open information.}
\label{fig:problem_model} \vspace{-0.25in}
\end{figure}
We consider the cellular network illustrated in
Fig.~\ref{fig:problem_model}. The network consists of $n$ nodes,
each of which has both open and private information to be
transmitted to a single base station over the associated uplink
channel. When a node is transmitting, every other node overhears the
transmission over the associated cross channel. We assume every
channel to be iid block fading, with a block size of $N_1$ channel
uses. The entire session lasts for $N_2$ blocks, which corresponds
to a total of $N=N_1N_2$ channel uses. We denote the instantaneous
achievable rate for the uplink channel of node $j$ by $R_j(k)$,
which is the maximum mutual information between output symbols of
node $j$ and received symbols at the base station over block $k$.
Likewise, we denote the rate of the cross channel between nodes $j$ and $i$ with $R_{ji}(k)$, which is the maximum mutual information between output symbols of node $j$ and input symbols of node $i$ over block $k$. Note that there is no actual data transmission between any pair of nodes, but parameter $R_{ji}(k)$ will be necessary, when we evaluate the private rates between node $j$ and the base station.

Even though our results are general for all channel state distributions, in numerical evaluations, we assume all channels to be {\em Gaussian} and the transmit power to be constant, identical to $P$ for all blocks $k,\ 1\leq k \leq N_2$. We represent the uplink channel for node $j$ and the cross channel between nodes $j$ and $i$ with a power gain (magnitude square of the channel gains) $h_j(k)$ and $h_{ji}(k)$ respectively over block $k$. We normalize the power gains such that the (additive Gaussian) noise has unit variance. Then, as $N_1 \rightarrow \infty$,
\vspace{-0.1in}
\begin{align}
\label{eq:uplink_rate}
R_j(k) &= \log(1+Ph_j(k)) \\
\label{eq:cross_channel_rate}
R_{ji}(k) &= \log(1+Ph_{ji}(k)) .
\vspace{-0.1in}
\end{align}
\textr{ Each node $j$ has a private and an open message, $W_j^\text{priv} \in \{1,\ldots , 2^{NR_j^{\text{priv}}}\}$ and $W_j^\text{open} \in \{1,\ldots , 2^{NR_j^{\text{open}}}\}$ respectively, to be transmitted to the base station over $N$ channel uses, where $R_j^{\text{priv}}$ and $R_j^{\text{open}}$ denote the (long-term) private and open information rates respectively, for node $j$.} Let the
vector of symbols received by node
$i$ be $\vc{Y}_i$. To achieve {\em perfect privacy}, following constraint must be satisfied by node $j$: for all $i\neq j$, \textr{
\begin{equation}
\label{eq:perfect_privacy}
\lim_{N\to \infty} \frac{1}{N}I(W_j^\text{priv};\vc{Y}_i) \leq \varepsilon
\end{equation}
for any given $\varepsilon > 0$.} We define the {\em instantaneous
private information rate} of node $j$ transmitted privately from
node $i$ over block $k$ as:
\begin{equation}
\label{eq:privacy_rate_2}
R_{ji}^p(k) = [R_j(k)-R_{ji}(k)]^+ ,
\end{equation}
where $[\cdot ]^+=\max(0,\cdot)$. \textr{It was shown in~\cite{Wyner} that rate~(\ref{eq:privacy_rate_2}) is achievable as $N_1\to \infty$ and~\cite{fading} took it a step further and showed that, as $N_1,N_2 \rightarrow \infty$, a long-term private information rate of $\E{R_{ji}^p(k)}$ is achievable.}

The amount of open traffic, $A_j^o(k)$, and private traffic,
$A_j^p(k)$, injected in the queues at node $j$ (shown in Fig.~\ref{fig:problem_model}) in block $k$ are both
selected by node $j$ at the beginning of each block. Open and
private information are stored in separate queues with sizes
$Q_j^o(k)$ and $Q_j^p(k)$ respectively. At any given block, a
scheduler chooses which node will transmit and the amount of open
and private information to be encoded over the block. We use the
indicator variable ${\cal I}_j(k)$ to represent the scheduler
decision:
\begin{equation}
{\cal I}_j(k)=\begin{cases} 1, & \text{private information from node $j$} \\ 0, & \text{otherwise} \end{cases}.
\end{equation}

When we evaluate the region of achievable open and private data rate
pairs for the single node scenario, in
Section~\ref{sec:ach_single_user}, we assume that the transmitting
node has perfect causal knowledge of its uplink channel and the
cross-channel at every block $k$. Thus, the achievable region of
private and open rates constitutes upper bound on the achievable
rates for each node, which we find subsequently for the multiuser
setting with partial CSI. \textr{For private opportunistic scheduler
in the multiuser setting, we assume that, each node $j$ has perfect
causal knowledge of the uplink channel rate, $R_j(k)$, and its prior
distribution. However, we assume that it only has the long-term
averages, $\E{R_{ji}(k)},\ i\neq j$ of its cross-channel rates. To
achieve privacy with this level of CSI, private opportunistic
scheduler uses an encoding scheme that encodes private information
over many packets.} When we formulate our problem as that of network
utility maximization problem, we only assume knowledge of
instantaneous channel gains {\em without requiring the knowledge of
prior distribution of channel gains}. Hence, private encoding is
performed over a single block length unlike the case with private
opportunistic scheduler.
Additionally, we analyze a more realistic scenario when the
instantaneous channel rates are not known \emph{perfectly}, but
estimated with some random additive error. The scheduled
transmitter, $j$, will encode at a rate
\[ \hat{R}_j^p(k) = [R_j(k)-\rho_j(k)]^+ , \]
where $\rho_j(k)$ is the rate margin, chosen such that the
estimation error is taken into account. Note that when
$\rho_j(k)<\max_{i\neq j} R_{ji}(k)$, then perfect privacy
constraint \eqref{eq:perfect_privacy} is violated over block $k$. In
such a case, we say that \emph{privacy outage} has occurred. The
probability of privacy outage over block $k$ when user $j$ is
scheduled, is represented as $p_j^{\text{out}}(\rho_j(k))$. Since
perfect privacy cannot be ensured over every block, we require that
expected probability of privacy outage of each user $j$ is below a
given threshold $\gamma_j$.

\ignore{Finally note that, even though, the main focus in this paper is the uplink
scenario, in Section~\ref{sec:achievability}, we generalize the
results for the private opportunistic scheduler to the downlink scenario as
well.}
\vspace{-0.1in}
\section{Achievable Rates and Private Opportunistic Scheduling}
\label{sec:achievability}

In this section, we evaluate the region of private and open rates
achievable by a scheduler for multiuser uplink and downlink setting.
We start with a single node transmitting, and thus, the scheduler
only chooses whether to encode private information at any given
point in time or not. We consider the possibility of both the
separate and the joint encoding of private and open data. For
multiuser transmission, we introduce our scheme, {\em private
opportunistic scheduling}, evaluate achievable rates and show that
it maximizes the sum private information rate achievable by any scheduler. Along with private opportunistic scheduling, we provide the associated
physical-layer private encoding scheme that encodes information over many blocks.
\noindent
\begin{figure}
\centerline{\includegraphics[height=0.8in]{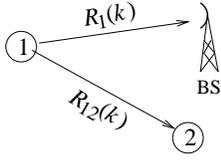}}
\caption{Single user private communication scenario.}
\label{fig:single_user}
\end{figure}
\vspace{-0.2cm}
\subsection{Single User Achievable Rates}
\label{sec:ach_single_user}
Consider the single user scenario in which the primary user (node 1) is transmitting information over the primary channel and a single secondary user (node 2) is overhearing the transmission over the secondary channel as shown in Fig.~\ref{fig:single_user}. In this scenario, we assume node 2 is passively listening without transmitting information and node 1 has perfect knowledge of instantaneous rates $R_1(k)$ and $R_{12}(k)$ for all $k$ as well as their sample distributions.
Over each block $k$, the primary user chooses the rate of private and open information to be transmitted to the intended receiver.
As discussed in~\cite{bc-secrecy} it is possible to encode open information at a rate $R_1(k)-R_{12}^p(k)$ over each block $k$,
jointly with the private information at rate $R_{12}^p(k)$. For that, one can simply replace the
randomization message of the binning strategy of the achievability
scheme with the open message, which is allowed to be decoded by the
secondary user. In the rest of the section, we analyze both the case in which open information can and cannot be encoded along with the private information. We find the region of achievable private and open information rates, $(R^{\text{priv}}_1,R^{\text{open}}_1)$, over the primary channel.

\subsubsection{Separate encoding of private and open messages}
\label{sec:separate_encoding}

First we assume that each block contains either private or open
information, but joint encoding over the same block is not allowed.
Recall that ${\cal I}_1(k)$ is the indicator variable, which takes
on a value $1$, if information is encoded privately over block $k$
and $0$ otherwise. Then, one can find $R^{\text{priv}}_1$,
associated with the point $R^{\text{open}}_1 = \alpha$ by solving
the following integer program: \vspace{-0.1in}
\begin{align}
\label{eq:sep_enc_obj} \max_{\{{\cal I}_1(k)\}\in\{0,1\}}  &\quad \E{{\cal I}_1(k)R_{12}^p(k)} \\
\label{eq:sep_enc_const} {\text{subject to}} &\quad
\E{(1-{\cal I}_1(k))R_1(k)} \geq \alpha,
\vspace{-0.1in}
\end{align}
where the expectations are over the joint distribution of the
instantaneous rates $R_1(k)$ and $R_{12}(k)$. Note that, since the
channel rates are iid, the solution, ${\cal I}_1^*(k)={\cal I}_1^*(R_1(k),R_{12}(k))$ will be a stationary policy. Also, a
necessary condition for the existence of a feasible solution is
$\E{R_1(k)} \geq \alpha$. Dropping the block
index $k$ for simplicity, the problem leads to the following
Lagrangian relaxation: \vspace{-0.1in}
\begin{multline}
\label{eq:sep_enc_lagrangian}
\min_{\lambda > 0}\ \max_{\{{\cal I}_1\}\in\{0,1\}} \E{{\cal I}_1R_{12}^p} +
\lambda \left( \E{(1-{\cal I}_1)R_1} - \alpha \right) \\
= \min_{\lambda > 0}\ \left\{ \max_{\{{\cal I}_1\}\in\{0,1\}} \int_0^{\infty} \int_0^{\infty}
\left[ {\cal I}_1R_{12}^p - \lambda (1-{\cal I}_1)R_1 \right] \right. \\ \left.
p(R_1,R_{12})\ dR_1dR_{12} -\lambda \alpha \right\} ,
\vspace{-0.1in}
\end{multline}
where $p(R_1,R_{12})$ is the joint pdf of $R_1$ and $R_{12}$.
For any given values of the Lagrange multiplier
$\lambda$ and $(R_1,R_{12})$ pair, the optimal policy will choose
${\cal I}_1^*(R_1,R_{12})=0$ if the integrant is maximized for ${\cal I}_1=0$, or it will choose ${\cal I}_1^*(R_1,R_{12})=1$ otherwise. If both
${\cal I}_1=0$ and ${\cal I}_1=1$ lead to an identical value, the policy
will choose one of them randomly. The solution can be summarized as
follows:
\begin{equation}
\label{eq:sep_enc_test}
\frac{R_{12}^p}{R_1} \test \lambda^*,
\end{equation}
where $\lambda^*$ is the value of $\lambda$ for which
$\E{(1-{\cal I}_1^*)R_1}=\alpha$, since $\lambda^* (\E{(1-{\cal I}_1)R_1} - \alpha) \leq 0$.

\begin{figure}
\centerline{\includegraphics[height=1.3in]{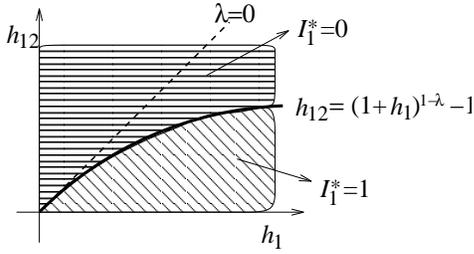}}
\caption{Optimal decision regions with separate encoding of private
and open messages.}
\label{fig:sep_enc_decision_regions}\vspace{-0.15in}
\end{figure}

For Gaussian uplink and cross channels described in Section~\ref{sec:model}, the solution can be obtained by plugging (\ref{eq:uplink_rate},\ref{eq:cross_channel_rate},\ref{eq:privacy_rate_2}) in (\ref{eq:sep_enc_test}): \textr{
\begin{equation}
\label{sep_enc_soln}
(1+Ph_1)^{1-\lambda^*} \test 1+Ph_{12} .
\end{equation}}
The associated solution ${\cal I}^*$ is graphically illustrated on
the $(h_1,h_{12})$ space in Fig.~\ref{fig:sep_enc_decision_regions} \textr{ for $P=1$}.
As the value of $\lambda$ varies between 0 and 1, the optimal
decision region for ${\cal I}=0$ increases from \textr{ the upper half of the first quadrant represented by $h_{12} \geq h_1$ to the
entire first quadrant, i.e., all $h_1,h_{12} \geq 0$.}
\noindent
\noindent
\begin{figure}
\centerline{\includegraphics[width=2.5in]{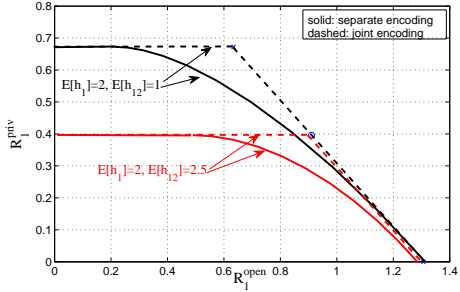}}
\caption{Achievable rate regions for the single user scenario with
iid Rayleigh block fading channels.}
\label{fig:single_user_ach_rates} \vspace{-0.25in}
\end{figure}
In Fig.~\ref{fig:single_user_ach_rates}, the achievable pair of
private and open information rates,
$(R^{\text{priv}}_1,R^{\text{open}}_1)$, is illustrated for iid
Rayleigh fading Gaussian channels, i.e., the power gains $h_1$ and
$h_{12}$ have an exponential distribution. We considered two
different scenarios in which the mean power gains,
$(\E{h_1},\E{h_{12}})$, are $(2,1)$ and $(2,2.5)$, and $P=1$. The
associated boundaries of the rate regions with separate encoding are
illustrated with solid curves. To plot these boundaries, we varied
$\lambda$ from 0 to 1 and calculated the achievable rate pair for
each point. Note that the flat portion on the top part of the rate
regions for separate encoding corresponds to the case in which
Constraint~(\ref{eq:sep_enc_const}) is inactive. It is also
interesting to note that as demonstrated in
Fig.~\ref{fig:single_user_ach_rates}, one can achieve non-zero
private information rates even when the mean cross channel gain
between node 1 and node 2 is higher than the mean uplink channel
gain of node 1.

\subsubsection{Joint encoding of private and open messages}
\label{sec:joint_encoding}

With the possibility of joint encoding of the open and private
information over the same block, the indicator variable ${\cal I}_1(k)=1$
implies that the private and open information rates are $R_{12}^p(k)$ and $R_1(k)-R_{12}^p(k)$ respectively over block $k$ simultaneously. Otherwise, i.e., if ${\cal I}_1(k)=0$, open encoding is used solely over the block.
To find achievable $R^{\text{priv}}_1$, associated with the point
$R^{\text{open}}_1 = \alpha$, one needs to consider a slightly different optimization problem this time:
\vspace{-0.1in}
\begin{align}
\label{eq:joint_enc_obj} \max_{\{{\cal I}_1(k)\}\in\{0,1\}} &\quad \E{{\cal I}_1(k)R_{12}^p(k)} \\
\label{eq:joint_enc_const} {\text{subject to}} &\ \E{(1-{\cal I}_1(k))R_1(k) + {\cal I}_1(k)(R_1(k)-R_{12}^p(k)) }\geq \alpha,
\vspace{-0.1in}
\end{align}

This optimization problem can be solved in a similar way by employing
Lagrangian relaxation as the problem considered in
Section~\ref{sec:separate_encoding}. \textr{ First, we specify two regions of
parameters for which the solution is trivial: 1) if $\E{R_1} < \alpha$,
no solution exists for~(\ref{eq:joint_enc_obj},\ref{eq:joint_enc_const}),
since the uplink channel capacity is not sufficient to meet the desired open rate, $\alpha$; 2) if $\E{R_1-R_{12}^p} > \alpha$, then ${\cal I}_1^*=1$ for all blocks, i.e., all open information will be encoded jointly with private information, since the remaining capacity over that is necessary to support private information is sufficient to serve open information at rate $\alpha$. In this case, Constraint~(\ref{eq:joint_enc_const}) is inactive and the achieved private information rate is $R^{\text{priv}}_1=\E{R_{12}^p}$.}

\textr{In all other cases, i.e., $\E{R_1-R_{12}^p} \leq \alpha \leq \E{R_1}$, it can be shown that the optimal solution can be achieved by the following probabilistic scheme\footnote{Note that the solution of Problem (\ref{eq:joint_enc_obj},\ref{eq:joint_enc_const}) is not unique and the described probabilistic solution is just one of them.}: For any given block,}
\begin{equation}
{\cal I}_1^*=\begin{cases} 1, & \text{w.p.}\ p^p \\
0, & \text{w.p.}\ 1-p^p
\end{cases} ,
\end{equation}
\textr{ independently of $R_1$ and $R_{12}$, where
$p^p = \frac{\E{R_1}-\alpha}{\E{R_{12}^p}}$. The details of the derivation of the described optimal scheme is given in~\cite{tech_report_Koksal_Ercetin}. With this solution, only a fraction $p^p$ of the blocks contain jointly encoded private and open information, and the remaining $1-p^p$ fraction of the blocks contain solely open information. Thus, for a given $\alpha$, the achieved private and open information rates can be found as $R^{\text{priv}}_1=p^p \E{R_{12}^p}=\E{R_1}-\alpha$ and $R^{\text{open}}_1=p^p \E{R_1-R_{12}^p}+(1-p^p)\E{R_1}=\alpha$ respectively. Rather surprisingly, it does not matter which blocks contain only open information and which ones contain jointly encoded private and open information, as long as the desired open information rate $\alpha$ is met. Consequently, a random scheme that chooses $1-p^p$ fraction of blocks for open information only and the rest for jointly encoded open and private information suffices to achieve the optimal solution.}

\textr{By the above analysis, one can conclude that the achievable rate region with joint encoding can be summarized by the intersection of two regions specified by: (i) $(R^{\text{priv}}_1+R^{\text{open}}_1) \leq \E{R_1}$ and (ii) $R^{\text{priv}}_1 \leq \E{R_{12}^p}$. Any point on the boundary of the region can be achieved by the simple probabilistic scheme described above. One can realize that this region is the maximum achievable rate region, since in our system, the total information rate (private and open) is upper bounded by the capacity, $\E{R_1}$, of uplink channel 1 and the achievable private rate is upper bounded by the secrecy capacity, $\E{R_{12}^p}$, of the associated wiretap channel. Thus, there exists no other scheme that can achieve a larger rate region than the one achieved by the simple probabilistic scheme.}


In Fig.~\ref{fig:single_user_ach_rates}, the achievable pairs of
private and open information rates,
$(R^{\text{priv}}_1,R^{\text{open}}_1)$ with joint encoding are
illustrated for the iid Rayleigh fading Gaussian channels with the
same parameters as the separate encoding scenario. The boundaries of
the regions are specified with dashed curves, which are plotted by
varying the value of $p^p$ from 0 to 1 and evaluating
$(\E{R_1-R_{12}^p},\E{R_{12}^p})$ pair for each value. Similar to
the separate encoding scenario, the flat portion on the top part of
the regions corresponds to the case in which
Constraint~(\ref{eq:joint_enc_const}) is inactive. \noindent
\begin{figure}
\centerline{\includegraphics[height=1in]{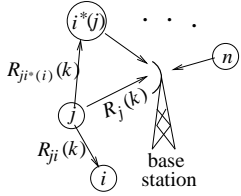}}
\caption{Multiuser private communication system - uplink}
\label{fig:multiuser_uplink}\vspace{-0.2in}
\end{figure}

\vspace{-0.15in}
\subsection{Private Opportunistic Scheduling and Multiuser Achievable Rates}
\label{sec:ach_multiuser}
\vspace{-0.05in}

In this section, we consider the multiuser setting described in
Fig.~\ref{fig:problem_model}. We \textr{introduce {\em private
opportunistic scheduling} (POS) for \ignore{both the downlink and}
the uplink scenario and prove that it achieves the maximum
achievable sum private information rate over the set of all
schedulers. POS schedules the node that has the largest
instantaneous private information rate, with respect to the ``best
eavesdropper'' node, which has the largest mean cross-channel rate.}
Each node ensures perfect privacy from its best eavesdropper node by
using a binning strategy, which requires only the average
cross-channel rates to encode the messages over many blocks.

\ignore{\subsubsection{Uplink Scenario}
\label{sec:multiuser_uplink}}

\ignore{First} We consider the multiuser uplink scenario given in
Fig.~\ref{fig:multiuser_uplink}.
\textr{We assume every node $j$ has perfect causal knowledge of its uplink channel rate, $R_j(k)$ for all blocks $k$ and the average cross-channel rates, $\E{R_{ji}(k)}$, for all $i\neq j$.}




\subsubsection{Private Opportunistic Scheduling for uplink}
\label{sec:multiuser_uplink}

We define the best eavesdropper of node $j$ as $i^*(j) \triangleq \argmax_{i\neq j} \E{R_{ji}(k)}$ and denote its average cross-channel rate with $\bar{R}_j^m \triangleq
\E{R_{ji^*(j)}(k)}$. \textr{Note that $i^*(j)$ does not change from one block to another.} In POS, only one of the nodes is scheduled for data transmission in any given block. In particular, in block $k$, we opportunistically schedule node
\[ j^M(k) \triangleq
\argmax_{j\in \{1,\ldots ,n\}} \left[ R_j(k) - \bar{R}_j^m \right] \]
if $\max_{j\in \{1,\ldots ,n\}} \left[ R_j(k) - \bar{R}_j^m \right] > 0$ and no node is scheduled for private information transmission otherwise, i.e., $j^M(k) = \emptyset$. \textr{In case of multiple nodes achieving the same maximum privacy rate, the tie can be broken at random.} Indicator variable ${\cal I}^{\text{POS}}_j(k)$ takes on a value $1$, if node $j$ is scheduled over block $k$ and $0$ otherwise.
We denote the probability that node $j$ be scheduled with $p^M_j \triangleq \Prob{j^M(k)=j}$ and the associated uplink channel rate when node $j$ is scheduled with $\bar{R}_j^M \triangleq \E{R_j(k)|j=j^M(k)}$, where the expectations are over the conditional joint distribution of the instantaneous rates of all uplink channels, given $j=j^M(k)$.

As will be shown shortly, private opportunistic scheduling achieves
a  private information rate $R_j^{\text{priv}}=p^M_j(\bar{R}_j^M -
\bar{R}_j^m)$ for all $j\in \{1,\ldots ,n\}$. To achieve this set of
rates, we use the following private encoding strategy based on
binning: To begin, node $j$ generates
$2^{Np^M_j(\bar{R}_j^M-\delta)}$ random binary sequences. Then, it
assigns each random binary sequence to one of
$2^{NR_j^{\text{priv}}}$ bins, so that each bin contains exactly
$2^{Np^M_j(\bar{R}_j^m-\delta)}$ binary sequences. We call the
sequences associated with a bin, the {\em randomization sequences}
of that bin. Each bin of node $j$ is one-to-one matched with a
private message $w \in \{1,\ldots ,2^{NR_j^{\text{priv}}}\}$
randomly. This selection \textr{ (along with the binary sequences
contained in each bin)} is revealed to the base station and all
nodes before the communication starts. Then, whenever the message to
be transmitted is selected by node $j$, the stochastic encoder of
that node chooses one of the randomization sequences associated with
each bin at random\footnote{\textr{In case of joint encoding of
private and open information, the randomization sequence is chosen
appropriately, corresponding to the desired open message.}},
independently and uniformly over all randomization sequences
associated with that bin. This particular randomization message is
used for the transmission of the message and is not revealed to any
of the nodes nor to the base station.

Private opportunistic scheduler schedules node $j^M(k)$ in each
block $k$ and the transmitter transmits $N_1 R_{j^M(k)}(k)$ bits of
the binary sequence associated with the message of node $j^M(k)$ for
all $k\in\{1,\ldots ,N_2\}$. Thus, asymptotically, the rate of data transmitted by node $j$ over $N_2$ blocks is identical to:
\begin{align}
\nonumber
\hspace{-0.07in} \lim_{N_1,N_2\rightarrow \infty} \frac{1}{N}
\sum_{k=1}^{N_2} N_1 {\cal I}^{\text{POS}}_j(k) R_j(k) &= \lim_{N_1,N_2\rightarrow
\infty} \frac{1}{N_2} \sum_{k=1}^{N_2} {\cal I}^{\text{POS}}_j(k) R_j(k) \\
&\geq
p^M_j(\bar{R}_j^M-\delta) \ \ \text{w.p.}\ 1
\label{eq:priv_opp_sched_1}
\end{align}
for any given $\delta > 0$ from strong law of large numbers. Hence, all
of $N(p^M_j(\bar{R}_j^M-\delta))$ bits, generated by each node $j$ is
transmitted with probability 1.


\subsubsection{Achievable uplink rates with private opportunistic scheduling}
\label{sec:POS_ach_rate}

\begin{theorem}
\label{the:pri_opp_sched_ach_up}
With private opportunistic scheduling, a private information rate of $R_j^{\text{priv}}=p^M_j(\bar{R}_j^M-\bar{R}_j^m)$ is achievable for each node $j$.
\end{theorem}

The proof of this theorem is based on an equivocation analysis and it can be found in Appendix~\ref{sec:appen_proof_0}. Next we show that private opportunistic scheduling maximizes the achievable sum private information rate among all schedulers.

\begin{theorem}
\label{the:pri_opp_sched_uplink}
Among the elements of the set of all schedulers, $\{{\cal I}(R_1,\ldots ,R_n)\}$, private opportunistic scheduler ${\cal I}^{\text{POS}}(R_1,\ldots ,R_n)$ maximizes the sum privacy uplink rate, $R_{\text{sum,up}}^{\text{priv}} = \sum_{j=1}^n R_j^{\text{priv}}$. Furthermore, the maximum achievable sum privacy uplink rate is
\[ R_{\text{sum,up}}^{\text{priv}}
= \sum_{j=1}^n \left[ p^M_j \left( \bar{R}_j^M - \bar{R}_j^m \right) \right]. \]
\end{theorem}

The proof of Theorem~\ref{the:pri_opp_sched_uplink} can be found in Appendix~\ref{sec:appen_proof_1}. There, we also show that the individual private information rates given in Theorem~\ref{the:pri_opp_sched_ach_up} are the maximum achievable individual rates with private opportunistic scheduling. Hence the converse of Theorem~\ref{the:pri_opp_sched_ach_up} also holds. Combining Theorems~\ref{the:pri_opp_sched_ach_up}~and~\ref{the:pri_opp_sched_uplink}, one can realize that private opportunistic scheduling achieves the maximum achievable sum private information rate. Thus, one cannot increase the individual private information rate a single node achieves with POS by an amount $\Delta > 0$, without reducing another node's private information rate by more than $\Delta$.

Next, we find the boundary of the region of achievable sum open and sum private uplink rate pair with {\em joint encoding of private and open information}. In opportunistic scheduling~\cite{Knopp:ICC:95,shroff} without any privacy constraint, the user with the best uplink channel is scheduled for all blocks $k$. Hence, the associated achievable rate can be written as $R_{\text{sum,up}}^{\text{opp}}=\E{\max_{j \in\{ 1,\ldots ,n \}}R_j(k)}$. Since this constitutes an upper bound for the achievable cumulative information rate~\cite{Knopp:ICC:95}, the total private and open information rate in our system cannot exceed $R_{\text{sum,up}}^{\text{opp}}$. Combining this with Theorem~\ref{the:pri_opp_sched_uplink}, we can characterize an outer bound for the achievable rate region for the sum rates as follows: (i) $R_{\text{sum,up}}^{\text{priv}} + R_{\text{sum,up}}^{\text{open}} \leq R_{\text{sum,up}}^{\text{opp}}$; (ii)
$R_{\text{sum,up}}^{\text{priv}} \leq \sum_{j=1}^{n} \left[ p^M_j \left(\bar{R}_j^M - \bar{R}_j^m \right) \right]$.
\textr{Next we illustrate this region and discuss how the entire region can be achieved by POS along with joint encoding of private and open messages.}

\noindent
\begin{figure}
\centerline{\includegraphics[height=1.8in]{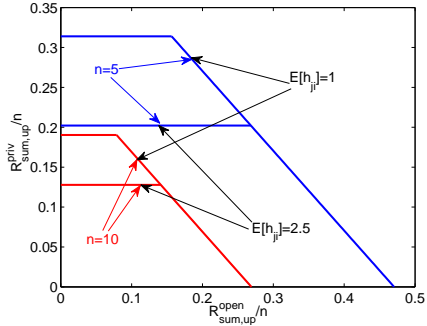}}
\caption{Bounds on the achievable sum rate region for the multiuser
uplink scenario with iid Rayleigh block fading channels.}
\label{fig:outer_bound_sum_rate}\vspace{-0.25in}
\end{figure}
The boundaries of this region is illustrated in
Fig.~\ref{fig:outer_bound_sum_rate} for a 5-node and a 10-node
system. We assume all channels to be iid Rayleigh fading with mean
uplink channel power gain $\E{h_j}=2$ and mean cross channel power
gain $\E{h_{ji}}=1$ or $2.5$ in two separate scenarios for all
$(i,j)$. Noise is additive Gaussian with unit variance and transmit
power $P=1$. In these graphs, sum rates are normalized with respect
to the number of nodes. One can observe that, the achievable sum
rate per node decreases from $0.31$ to $0.19$ bits/channel use/node
for $\E{h_{ji}}=1$ and from $0.2$ to $0.13$ bits/channel use/node
for $\E{h_{ji}}=2.5$ as the number of nodes increases from $5$ to
$10$. Also, the open rate per node drops from $0.47$ to $0.27$
bits/channel use/node with the same increase in the number of nodes.

\textr{Note that, any point on the part of the boundary specified by
(i) above (flat portion on the top part) is achievable by POS and
jointly encoding the private information with the appropriate amount
of open information used as a randomization message. For instance,
the corner point of two boundaries (intersection of (i) and (ii)) is
achieved when open information is used completely in place of
randomization messages by all nodes. All points on the part of the
boundary specified by (ii) can be achieved by time-sharing between
the corner point, and point
$\left(R_{\text{sum,up}}^{\text{opp}},0\right)$, which corresponds
to opportunistic scheduling (without privacy).}
\vspace{-0.15in}
%
\section{Dynamic Control of Private Communications}
\label{sec:control}

\textr{ In Section \ref{sec:achievability}, we determined the
achievable private information rate regions associated with private
opportunistic scheduling which encodes messages over {\em many
blocks}. Hence, the delay of decoding private information may be
extremely long. Also, the private opportunistic scheduler was based on the
availability of full CSI on the uplinks, and long-term average of
cross-channel rates. In this section, we investigate a dynamic
control algorithm which does not rely on any a priori knowledge of
distributions of direct- or cross-channel rates, and the private
information is encoded over {\em a single block}. Hence, a private
message can be decoded with a maximum delay of only a single block
duration.  Note that even though by encoding over many blocks one may
achieve higher private information rates, decoding delay may be a
more important concern in many practical scenarios. }

\textr{ In particular, each message $W_j^\text{priv}$ and
$W_j^\text{open}$ are broken into a sequence of messages,
$W_j^\text{priv}(k)$ and $W_j^\text{open}(k)$ respectively and each
element of the sequence is encoded into an individual packet,
encoded over block $k$. The {\em delay-limited} dynamic cross-layer
control algorithm opportunistically schedules the nodes with the
objective of maximizing the total expected utility gained from each
packet transmission while maintaining the stability of private and
open traffic queues.}
%
%
%
The algorithm takes as input the queue lengths and instantaneous
direct- and cross-channel rates, and gives as output the scheduled
node and its privacy encoding rate. In the sequel, we only consider
joint encoding of private and open information as described in
Section \ref{sec:joint_encoding}.

Let $g_j^p(k)$ and $g_j^o(k)$ be the utilities obtained by node $j$
from private and open transmissions over block $k$ respectively. Let
us define the instantaneous private information rate of node $j$ as $R_j^p(k)
\triangleq \min_{i\neq j}R_{ji}^p(k)$, where $R_{ji}^p(k)$ was
defined in (\ref{eq:privacy_rate_2}). Also, the instantaneous open
rate, $R_j^o(k)$, is the amount of open information node $j$
transmits over block $k$. The utility over block $k$ depends on
rates $R_j^p(k)$, and $R_j^o(k)$.  In general, this dependence can
be described as $g_j^p(k)=U^p_j(R_j^p(k))$ and
$g_j^o(k)=U^o_j(R_j^o(k))$. \textr{Assume that $U^p_j(0)=0$,
$U^o_j(0)=0$, and $U^p_j(\cdot)$, $U^0_j(\cdot)$ are concave
non-decreasing functions}.  We also assume that the utility of a
private transmission is higher than the utility of open transmission
at the same rate. The amount of open traffic $A_j^o(k)$, and private
traffic $A_j^p(k)$ injected in the queues at node $j$ have long term
arrival rates $\lambda_j^o$ and $\lambda_j^p$ respectively. Our
objective is to support a fraction of the traffic demand to achieve
a long term private and open throughput that maximizes the sum of
utilities of the nodes.

\vspace{-0.15in}
\subsection{Perfect Knowledge of Instantaneous CSI}
\vspace{-0.05in}

\textr{We first consider the case when every node $j$ has perfect
causal knowledge of its uplink channel rate, $R_j(k)$, and
cross-channel rates to all other nodes in the network $R_{ji}(k)$,
$\forall j\neq i$, for all blocks $k$. The dynamic control algorithm
developed for this case will then provide a basis for the algorithm
that we are going to develop for a more realistic case when
cross-channel rates are not known perfectly.} We aim to find the
solution of the following optimization problem: \vspace{-0.1in}
\begin{align}
&\max \sum_{j=1}^n \left(\E{g_j^p(k)}+\E{g_j^o(k)}\right) \label{eq:opt-objective-1}\\
& \text{subject to } (\lambda_j^o, \lambda_j^p)\in\Lambda \vspace{-0.1in}
\label{eq:const-stability-full-csi-2} 
\end{align}
The objective function in \eqref{eq:opt-objective-1} calculates the
total expected utility of open and private communications where
expectation is taken over the random achievable rates (random
channel conditions), and possibly over the randomized policy. The
constraint \eqref{eq:const-stability-full-csi-2} ensures that
private and open injection rates are within the achievable rate
region supported by the network denoted by $\Lambda$. In the
aforementioned optimization problem, it is implicitly required that
perfect secrecy condition given in \eqref{eq:perfect_privacy} is
satisfied in each block as $N_1\rightarrow \infty$.

The proposed cross-layer dynamic control algorithm is based on the
stochastic network optimization framework developed in \cite{Neely}.
This framework allows the solution of a long-term stochastic
optimization problem without requiring explicit characterization of
the achievable rate region, $\Lambda$.

We assume that there is an infinite backlog of data at the transport
layer of each node. Our proposed dynamic flow control algorithm
determines the amount of open and private traffic injected into the
queues at the network layer. The dynamics of private and open
traffic queues is given as follows:\vspace{-0.1in}
\begin{align}
Q_j^p(k+1)&=\left[ Q_j^p(k)-R_j^p(k)\right]^+ + A_j^p(k),
\label{secure_queue} \\
Q_j^o(k+1)&=\left[ Q_j^o(k)-R_j^o(k)\right]^+ + A_j^o(k), \vspace{-0.1in}
\label{unsecure_queue}
\end{align}
where $\left[x\right]^+=\max\{0,x\}$, and  \textr{the service rates
of private and open queues are given as, \vspace{-0.1in}
\begin{align}
R_j^p(k) &= {\cal I}_j^p(k)\left[R_j(k)-\max_{j\neq i} R_{ji}(k)
\right], \mbox{ and } \nonumber \\
R_j^o(k) &= {\cal I}_j^o(k)R_j(k)+ {\cal I}_j^p(k)(R_j(k)-R_j^p(k)). \vspace{-0.1in}
\nonumber
\end{align}
}where ${\cal I}_j^p(k)$ and ${\cal I}_j^o(k)$ are indicator
functions taking value ${\cal I}_j^p(k)=1$ when transmitting jointly
encoded private {\em and} open traffic, or ${\cal I}_j^o(k)=1$ when
transmitting {\em only} open traffic over block $k$ respectively.
Also note that at any block $k$, $\sum_j{\cal I}_j^p(k)+{\cal
I}_j^o(k)\leq 1$.

\noindent {\bf Control Algorithm:} The algorithm is a simple index policy and it executes the following steps in each block $k$:

\noindent {\bf (1) Flow control:} For some $V>0$, each
node $j$ injects $A_j^p(k)$ private and $A_j^o(k)$ open bits, where
\vspace{-0.1in}
\begin{align*}
\left( A_j^p(k),A_j^o(k) \right)=\argmax_{A^p,A^o}\ &\left\{V\left[
U_j^p(A^p)+U_j^o(A^o) \right]\right.\\
&\left.-\left( Q_j^p(k)A^p+Q_j^o(k)A^o \right)\right\} \vspace{-0.1in}
\end{align*}
\vspace{-0.1in}

\noindent {\bf (2) Scheduling:} Schedule node $j$ and
transmit jointly encoded {\em private  and open} traffic (${\cal I}_j^p=1$),
or {\em only open} (${\cal I}_j^o=1$) traffic, where
\vspace{-0.1in}
\[ ({\cal I}_j^p(k),{\cal I}_j^o(k))=\argmax_{{\cal I}^p,{\cal I}^o}\ \left\{ Q_j^p(k)R_j^p(k)+Q_j^o(k)R_j^o(k) \right\}, \vspace{-0.1in} \]
and for each node $j$, encode private data over each block $k$ at
rate
\vspace{-0.1in}
\[
R_j^p(k)={\cal I}_j^p(k)\left[ R_j(k)-\max_{i\neq j} R_{ji}(k)
\right], \] and transmit open data at rate
\[ R_j^o(k)= {\cal
I}_j^o(k)R_j(k)+ {\cal I}_j^p(k)(R_j(k)-R_j^p(k))
\]
\vspace{-0.08in}
\subsubsection*{Optimality of Control Algorithm} The optimality of the
algorithm can be shown using the Lyapunov optimization
theorem~\cite{Neely}. Before restating this theorem, we define the
following parameters.  Let $\mathbf{W}(k)=(W_1(k),\ldots, W_n(k))$
be the queue backlog process, and let our objective be the
maximization of time average of a scalar valued function $f(\cdot)$
of another process $\mathbf{R}(k)$ while keeping $\mathbf{W}(k)$
finite. Also define
$\Delta(\Wv(k))=\E{L(\mathbf{W}(k+1))-L(\mathbf{W}(k))|\mathbf{W}(k)}$
as the drift of some appropriate Lyapunov function $L(\cdot)$.

\begin{theorem}{(Lyapunov Optimization)~\cite{Neely}}
\label{thm:lyap} For the scalar valued function $f(\cdot)$, if there
exists positive constants $V$, $\epsilon$, $B$, such that for all
blocks $k$ and all unfinished work vector $\Wv(k)$ the Lyapunov
drift satisfies: \vspace{-0.1in}
\begin{equation}
\Delta(\Wv(k)) - V \E{f(\Rv(k))|\Wv(k)} \leq B - V f^* -\epsilon
\sum_{j=1}^{n} W_j(k), \vspace{-0.1in} \label{eq:optcond}
\end{equation}
then the time average utility and queue backlog satisfy:
\vspace{-0.05in}
\begin{align}
\liminf_{N_2\rightarrow\infty}\frac{1}{N_2}\sum_{k=0}^{N_2-1}\E{f(\Rv(k))} &\geq f^* - \frac{B}{V} \label{eq:LyapOpt1} \\
\limsup_{N_2\rightarrow\infty}\frac{1}{N_2}\sum_{k=0}^{N_2-1}
\sum_{j=1}^{n}\E{W_j(k)}
&\leq\frac{B+V(\bar{f}-f^*)}{\epsilon},\label{eq:LyapOpt2}
\vspace{-0.1in}
\end{align}
where $f^*$ is the \textr{maximal} value of $\E{f(\cdot)}$ and
$\bar{f} =
\limsup_{N_2\rightarrow\infty}\frac{1}{N_2}\sum_{k=0}^{N_2-1}\E{f(\Rv(k))}$.
\end{theorem}

For our purposes, we consider private and open unfinished work
vectors as $\mathbf{Q^p(k)}=(Q^p_1(k), Q^p_2(k),\ldots, Q^p_n(k))$,
and $\mathbf{Q^o(k)}=(Q^o_1(k), Q^o_2(k),\ldots, Q^o_n(k))$. Let
$L(\mathbf{Q^p},\mathbf{Q^o})$ be quadratic Lyapunov function of
private and open queue backlogs defined as: \textr{
\vspace{-0.1in}
\begin{equation}
L(\mathbf{Q^p(k)},\mathbf{Q^o(k)}) = \frac{1}{2}\sum_j
\left[(Q^p_j(k))^2+(Q^o_j(k))^2\right]. \label{eq:lyapunov-function}
\vspace{-0.1in}
\end{equation}}
Also consider the one-step expected Lyapunov drift, $\Delta(k)$ for
the Lyapunov function \eqref{eq:lyapunov-function} as:
\vspace{-0.1in}
\begin{multline}
\Delta(k) =
\mathbb{E}\left[ L(\mathbf{Q^p(k+1)},\mathbf{Q^o(k+1)}) \right. \\
- \left.
L(\mathbf{Q^p(k)},\mathbf{Q^o(k)})\bigm\vert\mathbf{Q^p(k)},\mathbf{Q^o(k)}\right].
\label{eq:lyapunov-drift}
\end{multline}
The following lemma provides an upper bound on $\Delta(k)$.

\begin{lemma}
\label{lemma:drift-1}
\begin{align}
\Delta(k)\leq B &- \sum_j \E{Q^p_j(k)(R^p_j(k)-A^p_j(k))\bigm\vert Q^p_j(k)} \nonumber  \\
&-\sum_j \E{Q^o_j(k)(R^o_j(k)-A^o_j(k))\bigm\vert Q^o_j(k)},
\label{eq:delta}
\end{align}
where $B>0$ is a constant.
\end{lemma}

The proof of Lemma \ref{lemma:drift-1} is given in Appendix
\ref{proof:drift-1}. Now, we present our main result showing that
our proposed dynamic control algorithm can achieve a performance
arbitrarily close to the optimal solution while keeping the queue
backlogs bounded.

\begin{theorem}
\label{thm:optimalcontrol-1}
 If $R_j(k)<\infty$ for all $j,k$, then dynamic control
algorithm satisfies:
\begin{align*}
\liminf_{N_2\rightarrow\infty}\frac{1}{N_2}\sum_{k=0}^{N_2-1}\sum_{j=1}^n
\E{g_j^p(k)+g_j^o(k)} &\geqslant g^* - \frac{B}{V} \\
\limsup_{N_2\rightarrow\infty}\frac{1}{N_2}\sum_{k=0}^{N_2-1}\sum_{j=1}^n
\E{Q_j^p(k)} &\leqslant \frac{B+V(\bar{g}-g^*)}{\epsilon_1} \\
\limsup_{N_2\rightarrow\infty}\frac{1}{N_2}\sum_{k=0}^{N_2-1}\sum_{j=1}^n
\E{Q_j^o(k)} &\leqslant \frac{B+V(\bar{g}-g^*)}{\epsilon_2} ,
\end{align*}
where $B,\epsilon_1,\epsilon_2>0$ are constants, $g^*$ is the
optimal solution of \eqref{eq:opt-objective-1}-\eqref{eq:const-stability-full-csi-2} and $\bar{g}$ is the maximum possible aggregate utility.
\end{theorem}

The proof of Theorem \ref{thm:optimalcontrol-1} is given in
Appendix~\ref{proof:optimalcontrol-1}.

\vspace{-0.15in}
\subsection{Imperfect Knowledge of Instantaneous CSI}
In the previous section, we performed our analysis assuming that at
every block {\em exact} instantaneous cross-channel rates are
available. However, unlike the uplink direct channel rate which can
be determined by the base station prior to the data transmission
(e.g., via pilot signal transmission), cross-channel rates are
harder to be estimated.
Indeed, in a non-cooperative network in which nodes do not exchange their
CSI, the cross-channel rates $\{ R_{ji}, j\neq i\}$ can only be
inferred by node $j$ from the received signals over the reverse
channel as nodes $j\neq i$ are transmitting to the base station.
Hence, at a given block, nodes only have {\em a posteriori} channel
distribution.  Based on this {\em a posteriori} channel
distribution, nodes may estimate CSI of their cross-channels.


Let us denote the {\em estimated} rate of the cross-channel $(j,i)$ with $\hat{R}_{ji}(k)$. We also define {\em cross-channel rate margin} $\rho_j(k)$ as the cross-channel rate a node uses when it encodes private information. More specifically, node $j$ encodes its private information at rate:
\begin{align}
R^p_j(k) = R_j (k) - \rho_j(k),
\end{align}
i.e., $\rho_j(k)$ is the rate of the randomization message node $j$ uses in the random binning scheme for privacy.
Note that, if
$\rho_j(k) < \max_{i\neq j}R_{ji}(k)$, then node $j$ will not meet the
perfect secrecy constraint at block $k$, leading to a {\em privacy outage}. In the event of a privacy outage, the privately encoded message is considered as an {\em open} message.
The probability of privacy outage over block $k$ for the scheduled node $j$, given the estimates of the cross channel rates is:
\begin{equation}
p_j^{\text{out}}(\rho_j(k)) = \Prob{\max_{i\neq
j}R_{ji}(k)>\rho_j(k)\Bigm\vert \{\hat{R}_{ji}(k),i\neq j\}}.
\end{equation}
Compare the aforementioned definition of {\em privacy outage} with
the {\em channel outage} \cite{Tse:Book:05} experienced in fast
varying wireless channels.  In time-varying wireless channels,
channel outage occurs when received signal and interference/noise
ratio drops below a threshold necessary for correct decoding of the
transmitted signal. Hence, the largest rate of reliable
communications at a given outage probability is an important measure
of channel quality.  In the following, we aim to determine utility
maximizing achievable privacy and open transmission rates for given
{\em privacy outage} probabilities.  In particular, we consider the
solution of the following optimization problem:
\begin{align}
&\max \sum_{j=1}^n \left(\E{g_j^p(k)}+\E{g_j^o(k)} \right) \label{eq:opt-objective-2}\\
& \text{subject to } (\lambda_j^o, \lambda_j^p)\in\Lambda,
\label{eq:const-stability-partial-csi}\\
& \text{and } p_j^{\text{out}}(\rho_j(k))= \gamma_j,
\label{eq:const-secrecy-outage}
\end{align}
where $\gamma_j$ is the tolerable privacy outage probability.
Aforementioned optimization problem is the same as the one given for
perfect CSI except for the last constraint.
The additional constraint \eqref{eq:const-secrecy-outage} requires
that only a certain prescribed proportion of private transmissions
are allowed to violate the perfect privacy constraint. Due to
privacy outages we define {\em private goodput} of user $j$ as
$\E{R_j^p(k)\left( 1-p_j^{\text{out}}(\rho_j(k))\right)}$. Note that
private goodput only includes private messages for which perfect
privacy constraint is satisfied.  All private messages for which
\eqref{eq:perfect_privacy} is violated are counted as successful
open transmissions.


Similar to the perfect CSI case, we argue that a dynamic policy can
be used to achieve asymptotically optimal solution. Unlike the
algorithm given in the perfect CSI case, the algorithm for imperfect
CSI first determines the private data encoding rate so that the
privacy outage constraint \eqref{eq:const-secrecy-outage} is
satisfied in current block. Hence, the private encoding rate at a
particular block is determined by the estimated channel rates and
the privacy outage constraint.


\noindent {\bf Control Algorithm:} Similar to the perfect CSI case, our algorithm involves two steps in each block $k$:
  
\noindent {\bf (1) Flow Control:} For some $V>0$, each node
injects $A_j^p(k)$ private and $A_j^o(k)$ open bits, where \vspace{-0.1in}
\begin{align}
 \left(A_j^p(k), A_j^o(k)\right)=&\argmax_{A_j^p, A_j^o}\,\, V\left[U_j^p(A_j^p)(1-\gamma_j)\right.\nonumber\\
 & \left.+U_j^o(A_j^o)(1-\gamma_j)+U_j^o(A_j^o+A_j^p)\gamma_j\right]\nonumber \\
 &  -Q_j^p(k)A_j^p-Q_j^o(k)A_j^o . \vspace{-0.15in} \label{eq:flowcontrol}
\end{align}

\noindent {\bf (2) Scheduling:} Schedule node $j$ and transmit jointly encoded
{\em private  and open} traffic (${\cal I}_j^p=1$) or {\em only} open (${\cal
I}_j^o=1$) traffic, where \vspace{-0.1in}
\begin{align*}
\left({\cal I}_j^p(k),{\cal I}_j^o(k)\right)=\argmax_{{\cal
I}^p,{\cal I}^o}\ &\left\{
Q_j^p(k)R_j^p(k)+Q_j^o(k)R_j^o(k)\right\}. \vspace{-0.1in}
\end{align*}
For each node $j$, encode private data over each
block $k$ at rate \vspace{-0.15in}
\[ R_j^p(k) = {\cal I}_j^p(k)\left[ R_j(k)-\rho_j(k) \right], \quad
\rho_j(k) = p_j^{{\text{out}}^{-1}}(\gamma_j), \vspace{-0.1in} \]
and transmit open data at rate \vspace{-0.1in}
\[ R_j^o(k)= {\cal
I}_j^o(k)R_j(k)+ {\cal I}_j^p(k)(R_j(k)-R_j^p(k)). \vspace{-0.1in}
\]
\vspace{-0.15in}
\subsubsection*{Optimality of Control Algorithm}
The optimality of the control algorithm with imperfect CSI can be
shown in a similar fashion as for the control algorithm with perfect
CSI.  We use the same Lyapunov function defined in
\eqref{eq:lyapunov-function} which results in the same one-step
Lyapunov drift function \eqref{eq:lyapunov-drift}.  Hence, Lemma
\ref{lemma:drift-1} also holds for the case of imperfect CSI, but
with a different constant $B^\prime$ due to the fact that higher
maximum private information rates can be achieved by allowing privacy outages.

Lyapunov Optimization Theorem suggests that a good control strategy
is the one that minimizes the following: \vspace{-0.1in}
\begin{equation}
\Delta^U(k)=\Delta(k) - V \E{\sum_j (g^p_j(k)+g^o_j(k))\Bigm\vert
\mathbf{Q^p}(k),\mathbf{Q^o}(k)} \vspace{-0.1in} \label{eq:deltawithreward-2}
\end{equation}
In \eqref{eq:deltawithreward-2}, expectation is over all possible
channel states. The expected utility for private and open
transmissions are respectively given as: \vspace{-0.1in}
\begin{align}
\E{g_j^p(k)}&= \E{g_j^p(k)|{\cal I}_j^p(k),\rho_j(k)} \nonumber \\
&=(1-\gamma_j)\E{U^p_j\left(A_j^p(k)\right)}, \label{eq:exp-util-private}\\
\E{g_j^o(k)} &= \E{g_j^o(k)|{\cal I}_j^p(k),{\cal I}_j^o(k), \rho_j(k)} \nonumber\\
&=\gamma_j\E{U^o_j(A_j^p(k)+A_j^o(k))} \nonumber \\
&\hspace{1.5cm} +(1-\gamma_j)
\E{U^o_j(A_j^o(k))}  \vspace{-0.1in} \label{eq:exp-util-open}.
\end{align}

Note that \eqref{eq:exp-util-private}-\eqref{eq:exp-util-open} are
obtained due to Constraint \eqref{eq:const-secrecy-outage}. By
combining Lemma \ref{lemma:drift-1} with
\eqref{eq:exp-util-private}-\eqref{eq:exp-util-open} we may obtain
an upper bound for \eqref{eq:deltawithreward-2}, as follows: \vspace{-0.1in}
\begin{align}
\Delta^U(k)< &B^\prime-\sum_j \mathbb{E}\left[
Q^p_j(k)[R^p_j(k)-A^p_j(k)]\right]\nonumber \\
-\sum_j \mathbb{E}&\left[ Q^o_j(k)[R^o_j(k)-A^o_j(k)]\right] -V \mathbb{E}\left[ \sum_j (1-\gamma_j) U_j^p\left(A_j^p(k)\right)\right.\nonumber\\
&\hspace{-0.4cm}\left.+ \sum_j \gamma_j U_j^o(A_j^p(k)+A_j^o(k)) +
(1-\gamma_j)U_j^o(A_j^o(k))\right]. \vspace{-0.1in} \label{drift_final-2}
\end{align}
Now, it is clear that the proposed dynamic control algorithm
minimizes the right hand side of \eqref{drift_final-2}.  The steps
of proving the optimality of the dynamic control algorithm are
exactly the same as those given in Theorem
\ref{thm:optimalcontrol-1}, and hence, we skip the details.

\begin{theorem}
If $R_j(k)<\infty$, for all $j,k$ then dynamic control algorithm satisfies:
\begin{align}
\liminf_{N_2\rightarrow\infty}\frac{1}{N_2}\sum_{k=0}^{N_2-1}\sum_{j=1}^n{\mathbb{E}}[g_j^p(k)+g_j^o(k)] &\geq {g^\prime}^*- \frac{B^\prime}{V}  \\
\limsup_{N_2\rightarrow\infty}\frac{1}{N_2}\sum_{k=0}^{N_2-1}\sum_{j=1}^n{\mathbb{E}}[Q_j^p(k)]&\leq\frac{B^\prime+V(\bar{{g^\prime}}-{g^\prime}^*)}{\epsilon_2^\prime}\nonumber\\
\limsup_{N_2\rightarrow\infty}\frac{1}{N_2}\sum_{k=0}^{N_2-1}\sum_{j=1}^n{\mathbb{E}}[Q_j^o(k)]&\leq\frac{B^\prime+V(\bar{{g^\prime}}-{g^\prime}^*)}{\epsilon_1^\prime},\nonumber
\end{align}
where $B^\prime,\epsilon_1^\prime,\epsilon_2^\prime>0$ are
constants, ${g^\prime}^*$ is the optimal solution of
\eqref{eq:opt-objective-2}-\eqref{eq:const-secrecy-outage} and
$\bar{{g^\prime}}$ is the maximum possible aggregate utility.
\end{theorem}

\vspace{-0.1in}
\section{Numerical Results}

In our numerical experiments, we considered a network consisting of
ten nodes and a single base station. The direct channel between a
node and the base station, and the cross-channels between pairs of
nodes are modeled as iid Rayleigh fading Gaussian channels. Thus,
direct-channel and cross-channel power gains are exponentially
distributed with means chosen uniformly randomly in the intervals
$[2,8]$,  and $[0,1]$, respectively.  The noise normalized power is
$P=1$. In our simulations, we consider both of the cases when
perfect instantaneous CSI is available, and when instantaneous CSI
can only be estimated with some error. Unless otherwise indicated,
in the case of imperfect CSI, we take the tolerable privacy outage
probability as $0.1$.
We assumed the use of an unbiased estimator for the cross-channel power gains and modeled the associated estimation error with a Gaussian random variable:
\[{\hat h}_{ji}(k)=h_{ji}(k)+e_{ji}(k),\]
where $e_{ji}(k) \sim {\cal N}(0,\sigma^2)$ for all $k$. Gaussian
estimation error can be justified as discussed in~\cite{Frenger} or
by the use of a recursive ML estimator as in~\cite{Kok_Sch_10}.
Unless otherwise stated, we take $\sigma=0.5$, i.e., the estimation
error is rather significant relative to the mean cross-channel gain.
Note that, in this section, we choose the margin $\rho_j(k)$ such
that
\[ \Prob{\rho_j(k) < \max_{i\neq j}\left[ \log(1+Ph_{ji})\right] \biggm\vert \{\hat{h}_{ji},i\neq j \}} \leq \gamma_j .\]
We consider logarithmic private and open utility functions where the
private utility is $\kappa$ times more than open utility at the same
rate. More specifically, we take for a scheduled node $j$,
$U_j^p(k)=\kappa\cdot\log(1+R_j^p(k))$, and
$U_j^o(k)=\log(1+R_j^o(k))$.  We take $\kappa=5$ in all the
experiments except for the one inspecting the effect of $\kappa$.
The rates depicted in the graphs are per node arrival or service
rates calculated as the total arrival or service rates achieved by
the network divided by the number of nodes, i.e., the unit of the
plotted rates is bits/channel use/node. Finally, for perfect CSI, we
only plot the service rates since arrival and service rates are
identical.

\begin{figure*}
\centerline{
\subfloat[Long-term Utility]{\includegraphics[width=3in]{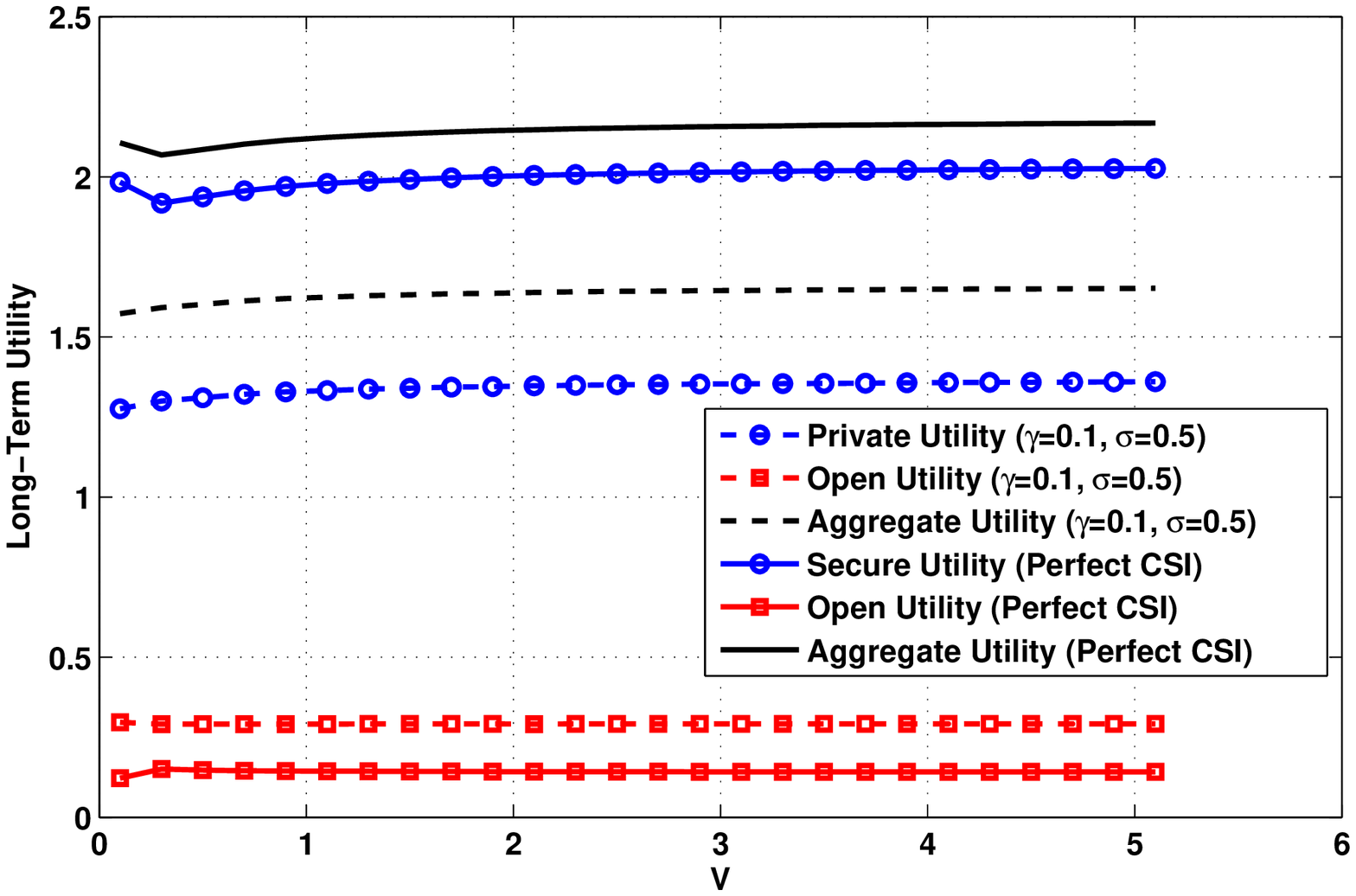}%
\label{fig:V_utility}} \hfil
\subfloat[Average queue length]{\includegraphics[width=3in]{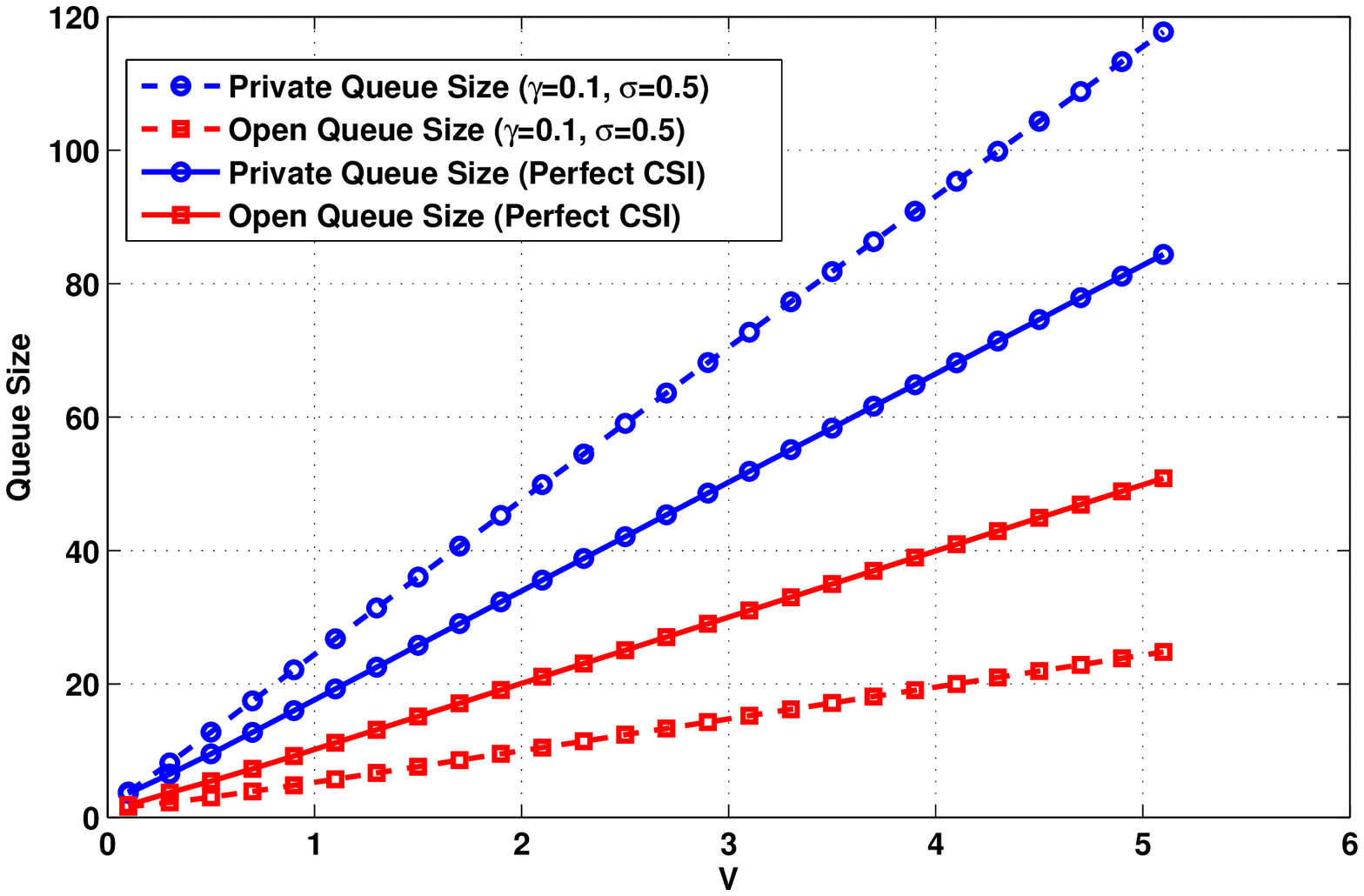}%
\label{fig:V_queue}} } \caption{Numerical results with respect to
optimization parameter $V$.}  \vspace{-0.2in}
\end{figure*}

In Fig.~\ref{fig:V_utility}-\ref{fig:V_queue}, we investigate the
effect of system parameter $V$ in our dynamic control algorithm.
Fig.~\ref{fig:V_utility} shows that for $V>4$, long-term utilities
converge to their optimal values fairly closely.  It is also
observed that CSI estimation error results in a reduction of
approximately 25\% in aggregate utility.  Fig.~\ref{fig:V_queue}
depicts the well-known relationship between $V$ and queue backlogs,
where queue backlogs increase when $V$ is increased.

\begin{figure*}
\centerline{
\subfloat[Perfect CSI]{\includegraphics[width=3in]{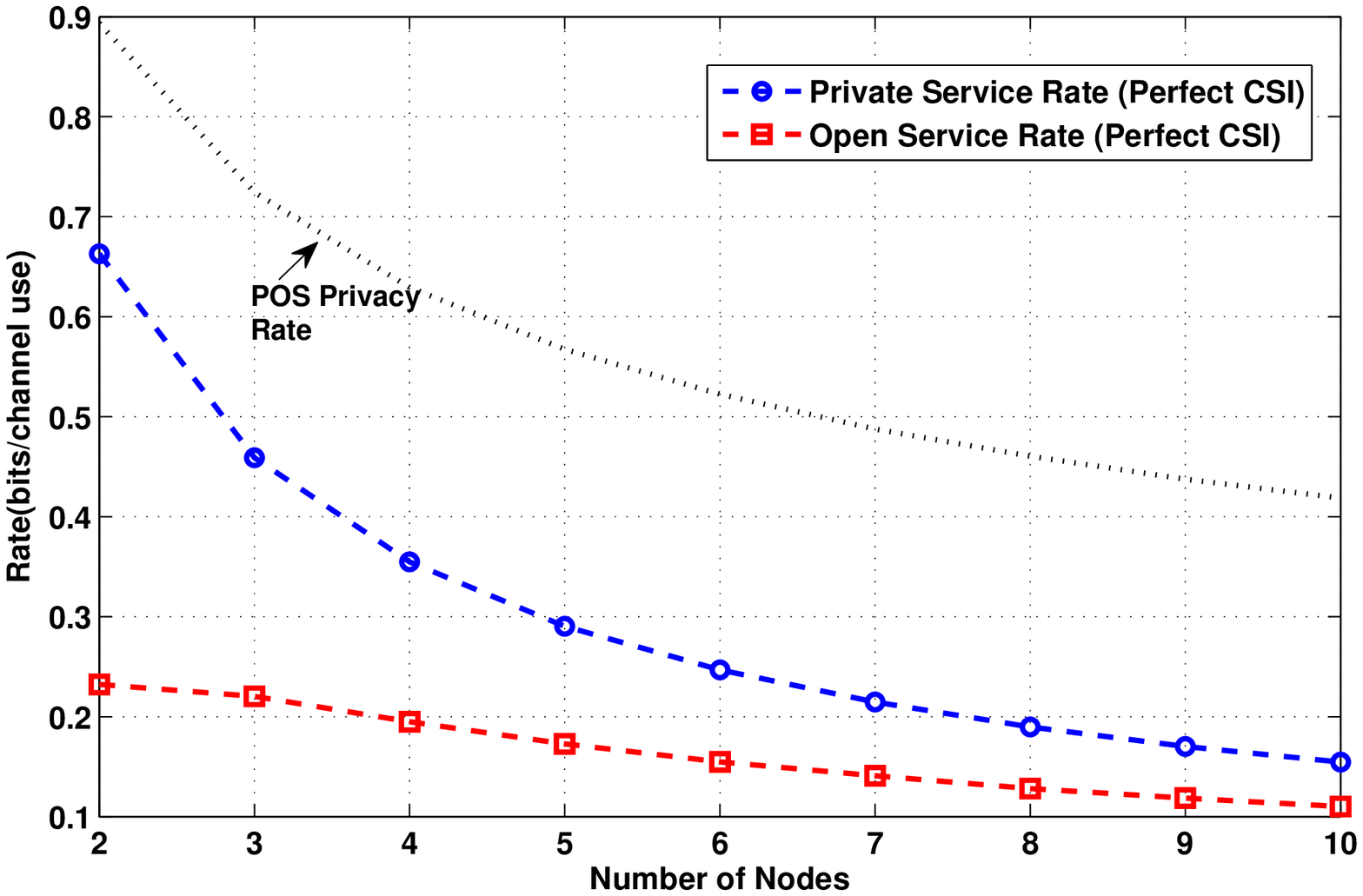}%
\label{fig:Node_perfect}} \hfil
\subfloat[Imperfect CSI]{\includegraphics[width=3in]{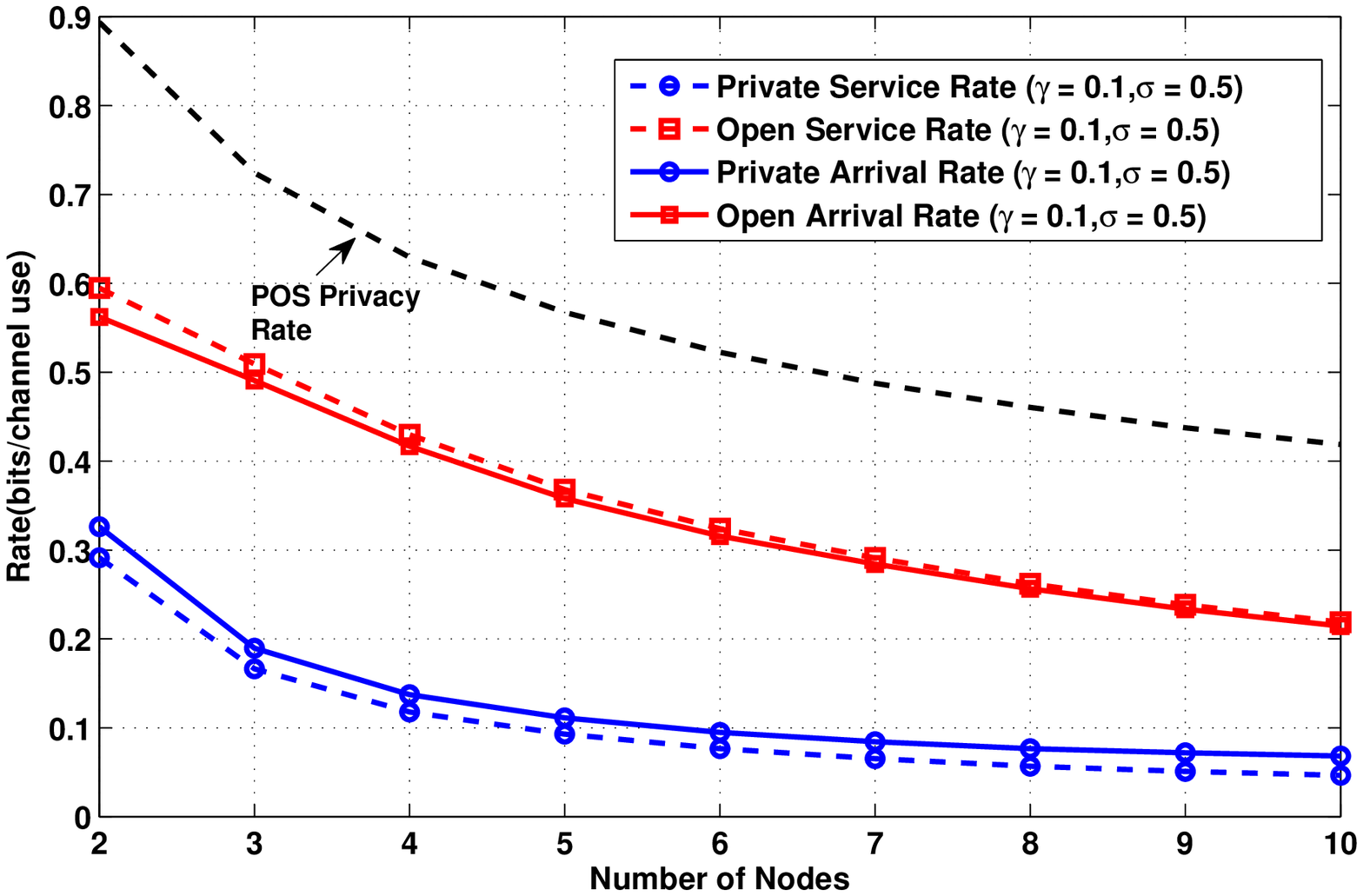}%
\label{fig:Node_partial}}} \caption{Private and open rates with
respect to number of nodes} \vspace{-0.2in}
\end{figure*}

In Fig.~\ref{fig:Node_perfect}-\ref{fig:Node_partial}, the effect of
increasing number of nodes on the achievable private and open rates
obtained with the proposed dynamic control algorithm is shown. In
both figures, the private information rate achieved by POS
algorithm given in Section \ref{sec:achievability} is also depicted.
From Fig.~\ref{fig:Node_perfect}, we first notice that by using the
dynamic control algorithm which is based only on the instantaneous
CSI, the private service rate is reduced by more than 25\% as
compared to the maximum private information rate achieved by POS which uses a priori CSI to encode over many blocks.
This difference increases with increasing number of nodes. However, for
both POS and dynamic control algorithms, the achievable rates
decrease with increasing number of nodes since more nodes overhear
ongoing transmissions. Meanwhile, open service rate also decreases
due to the fact that there is a smaller number of transmission
opportunities per node with increasing number of nodes.
Fig.~\ref{fig:Node_partial} depicts that private service rate has
decreased by approximately 50\% due to CSI estimation errors. It is
also interesting to note that private arrival rate is higher than
the private service rate, since all private messages for which
perfect privacy constraint cannot be satisfied are considered as
successful open messages. Hence, open service rate is observed to be
higher than the open arrival rate.

\begin{figure*}
\centerline{
\subfloat[Perfect CSI]{\includegraphics[width=3in]{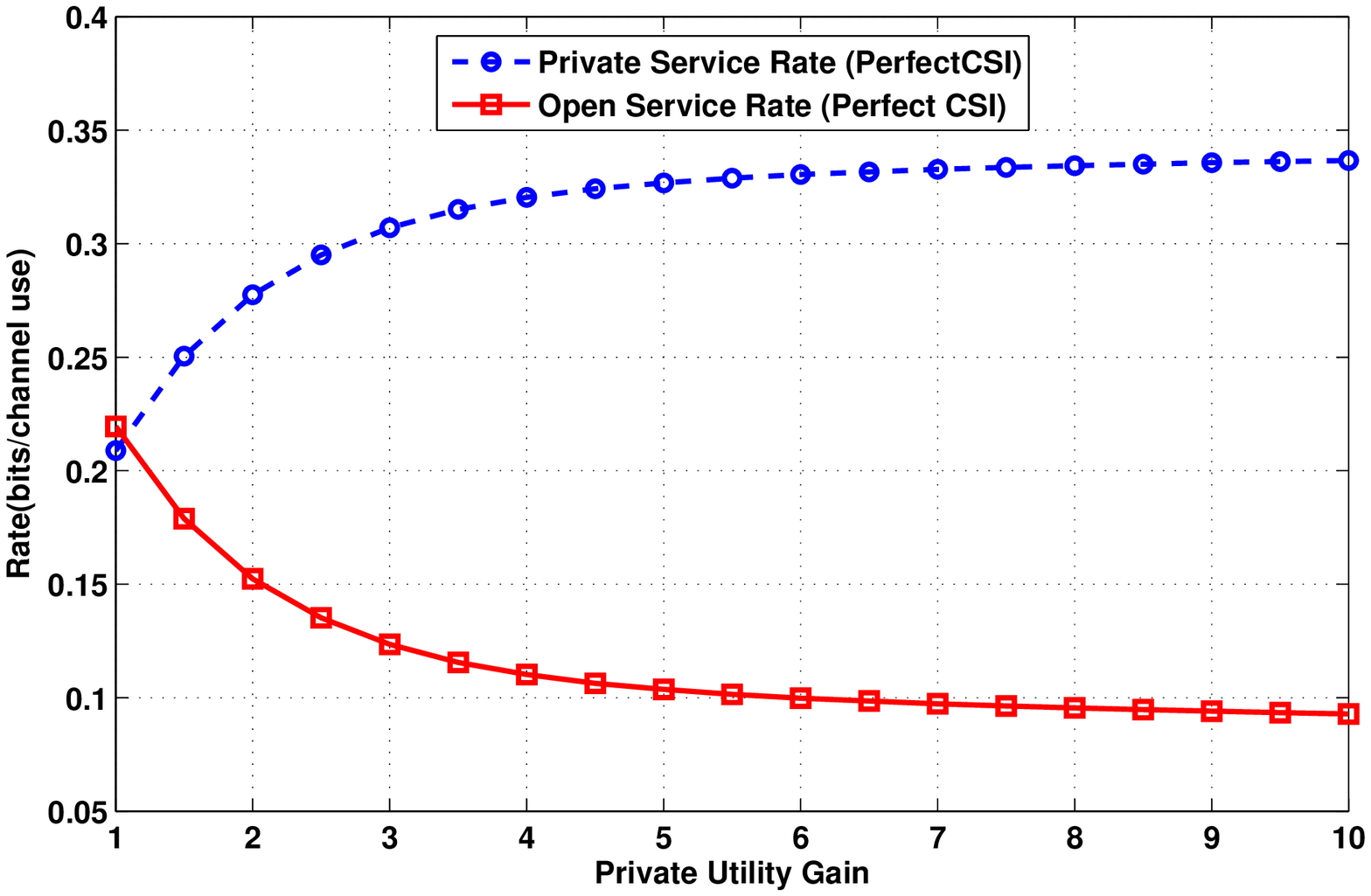}%
\label{fig:Gain_perfect}}\hfil
\subfloat[Imperfect CSI]{\includegraphics[width=3in]{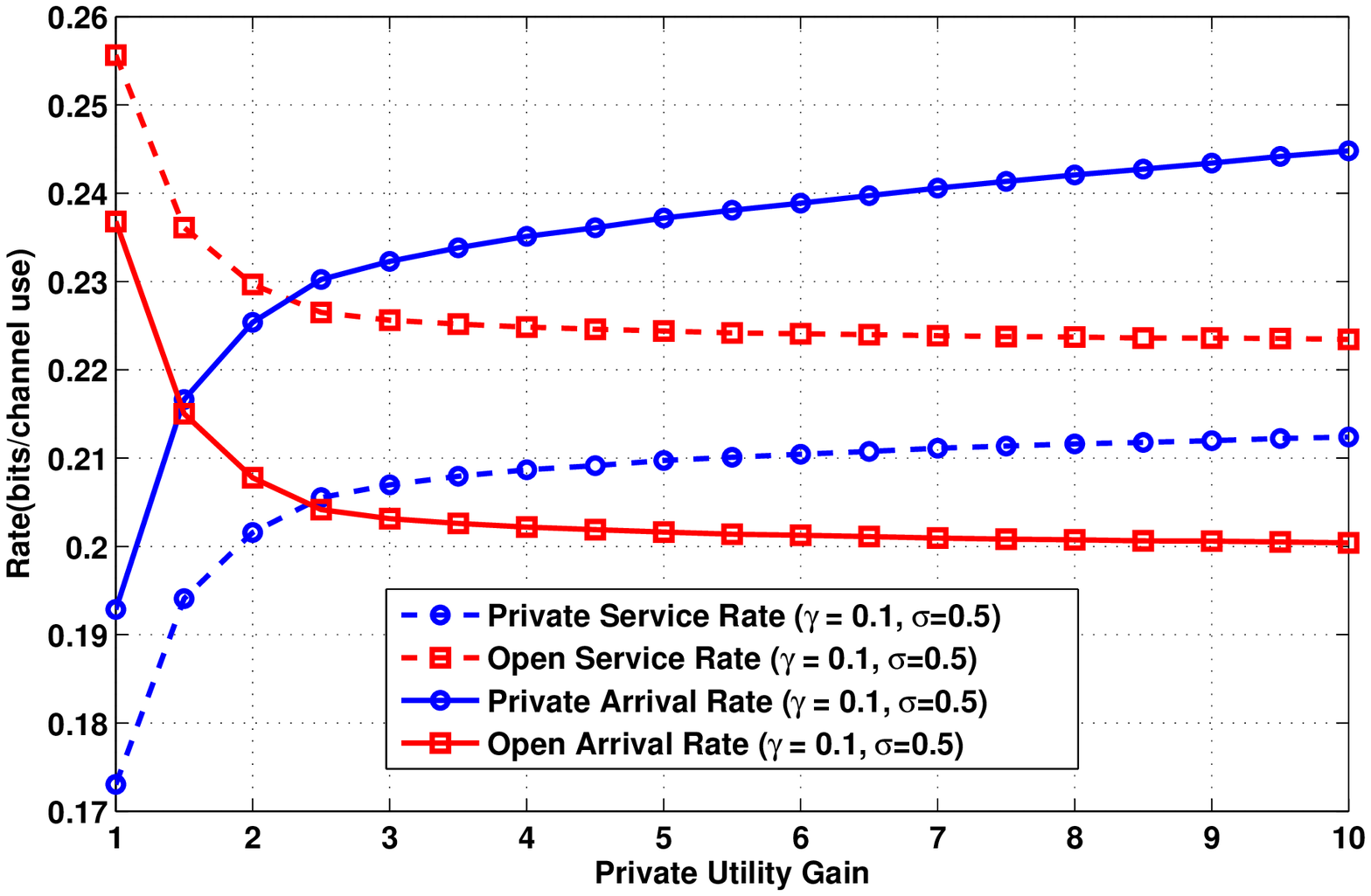}%
\label{fig:Gain_partial}}  } \caption{Private and open rates with
respect to increasing amount of private utility gain.}
\vspace{-0.2in}
\end{figure*}

We next analyze the effect of $\kappa$, which can also be
interpreted as the ratio of utility of private and open
transmissions taking place at the same rate. We call this ratio
{\em private utility gain}. Fig.~\ref{fig:Gain_perfect} shows that when
private utility gain is greater than 5, then the private and open
service rates converge to their respective limits. These limits
depend on the channel characteristics, and their sum is
approximately equal to the maximum achievable rate of the channel. However, when
there is CSI estimation error, Fig.~\ref{fig:Gain_partial} shows
that although an identical qualitative relationship between arrival
rates and private utility gain is still observed, private service
rate is lower than the private arrival rate by a fraction of $\gamma$
almost uniformly in the range of $\kappa$.

\begin{figure*}
\centerline{ \subfloat[Effect of $\gamma$.]{\includegraphics[width=3in]{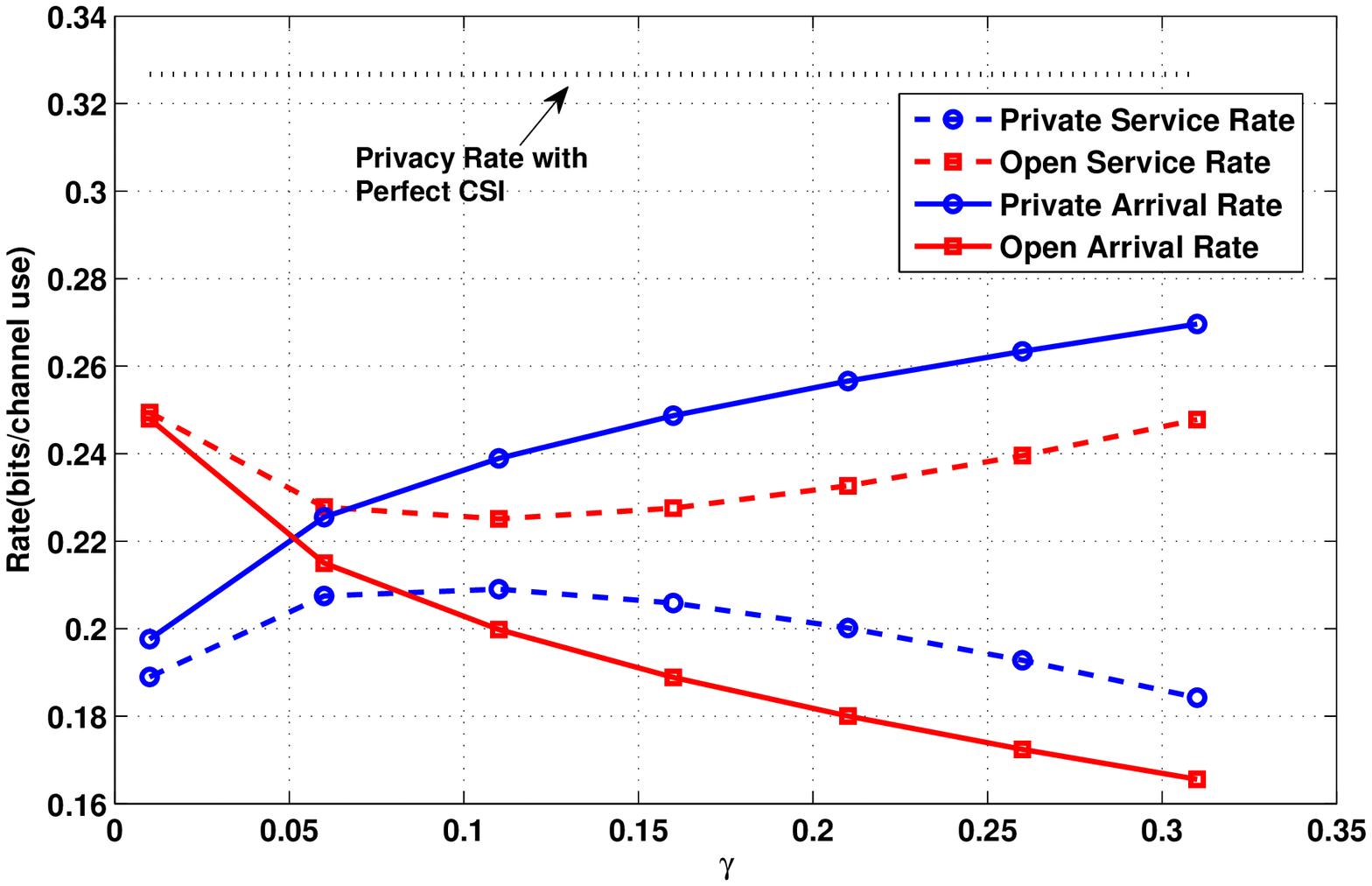}%
\label{fig:gamma}}\hfil \subfloat[Effect of $\sigma$.]{\includegraphics[width=3in]{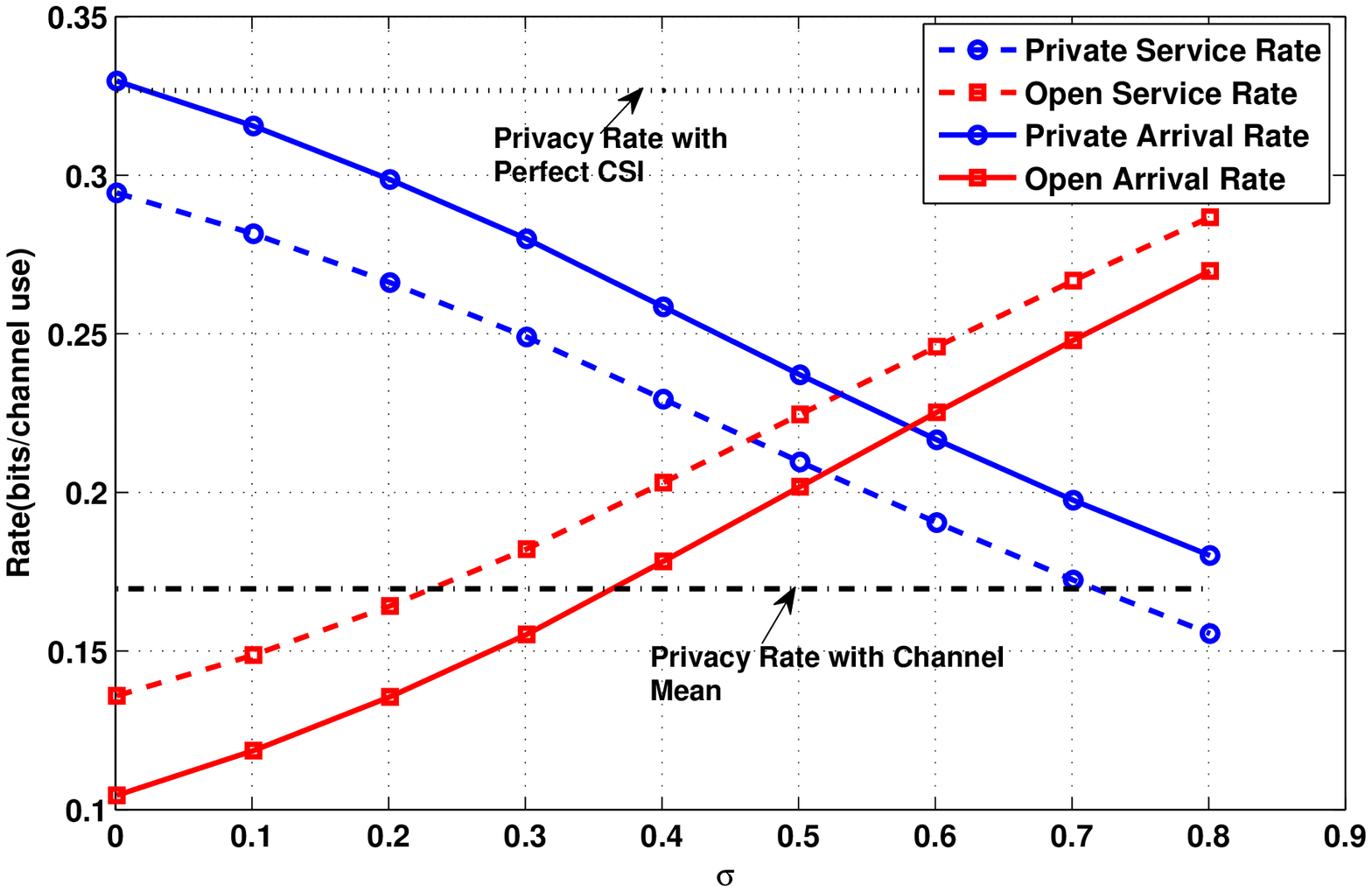}%
\label{fig:sigma}}  } \caption{Private and open rates with respect
to tolerable privacy outage probability.} \vspace{-0.2in}
\end{figure*}

In Fig.~\ref{fig:gamma}, we investigate the effect of the tolerable
privacy outage probability. It is interesting to note that private
service rate increases initially with increasing tolerable outage
probability. This is because for low $\gamma$ values, in order to
satisfy the tight privacy outage constraint, a low instantaneous
private information rate is chosen. However, when $\gamma$ is high more privacy
outages are experienced at the expense of higher instantaneous
private information rates. This is also the reason why we observe that the
difference between the private service and arrival rates is
increasing. We note that when CSI estimation error is present, the
highest private service rate is obtained when $\gamma$ is
approximately equal to $0.1$.  The highest private service rate with
CSI estimation error is approximately 30\% lower than the private
service rate with the perfect CSI.

We finally investigate the effect of the quality of CSI estimator in
Fig.~\ref{fig:sigma}. For this purpose, we vary the standard
deviation of the Gaussian random variable modeling the estimation
error.  As expected the highest private service rate is obtained
when $\sigma=0$. However, it is important to note that this value is
still lower than the private service rate with perfect CSI, since
privacy outages are still permitted in $10$\% of private
transmissions. We have also investigated the performance of the
dynamic control algorithm when a posteriori CSI distribution is not
available. In this case, scheduling and flow control decisions are
based only on the mean cross channel gains. When only mean cross
channel gains are available, the achieved private service rate per
node is approximately equal to $0.16$ bits per channel use, which is
significantly lower than the private service rate with perfect CSI.
In particular, it is only when the standard deviation of the
estimation error is $0.7$ that the private service rate with noisy
channel estimator has the same private service rate achievable
utilizing only mean channel gains.

%

\vspace{-0.2in}
\section{Conclusions}
In this paper, we studied the achievable private and open information rate regions of single- and multi-user wireless networks with node scheduling. We introduce private opportunistic scheduling along with a private encoding strategy, and show that it maximizes the sum private information rate for both multiuser uplink communication when perfect CSI is available for only the main uplink channels. Then, we described a cross-layer dynamic algorithm that works without prior distribution of channel states. We prove that our algorithm, which is based on simple index policies, achieves utility arbitrarily close to achievable optimal utility. The simulation results also verify the efficacy of the algorithm.

As a future direction, we will investigate the cooperation among nodes, e.g., intelligent jamming from cooperating nodes, as a means to improve the achievable private information rates.  We will also investigate
an extension of the dynamic control policy for imperfect CSI, where
the optimal privacy outage probability is also determined by the
algorithm.

\vspace{-0.1in}
\vspace{-0.0in}
\bibliographystyle{ieeetr}
\bibliography{macros,theo09,ent,paper_chan,tcom08,net,books,comm,macros_abbrev,SerdarSectionBib,sensornetwork,secrecy,scaling_laws,References,infocom10,new,newsecurity,Ref-OE}

\newcommand{\noopsort}[1]{} \newcommand{\printfirst}[2]{#1}
  \newcommand{\singleletter}[1]{#1} \newcommand{\switchargs}[2]{#2#1}
\begin{thebibliography}{10}

\bibitem{fading}
P.~K. Gopala, L.~Lai, and H.~{El Gamal}, ``On the secrecy capacity of fading
  channels,'' {\em IEEE Trans. Inform. Theory}, vol.~54, pp.~4687--4698, Oct.
  2008.

\bibitem{Barros:ISIT:06}
J.~Barros and M.~R.~D. Rodrigues, ``Secrecy capacity of wireless channels,'' in
  {\em Proc. IEEE Int. Symposium Inform. Theory}, (Seattle, WA), pp.~356--360,
  July 2006.

\bibitem{Liang:TIT:061}
Y.~Liang, H.~V. Poor, and {S. Shamai (Shitz)}, ``Secure communication over
  fading channels,'' {\em IEEE Trans. Inform. Theory}, vol.~54, pp.~2470--2492,
  June 2008.

\bibitem{Lai:TIT:07}
L.~Lai, H.~{El Gamal}, and H.~V. Poor, ``The wiretap channel with feedback:
  Encryption over the channel,'' {\em IEEE Trans. Inform. Theory}, vol.~54,
  pp.~5059 -- 5067, Nov. 2008.

\bibitem{Ardestanizadeh:TIT:09}
E.~Ardestanizadeh, M.~Franceschetti, T.~Javidi, and Y.~Kim, ``The secrecy
  capacity of the wiretap channel with rate-limited feedback,'' {\em IEEE
  Trans. Inform. Theory}, 2009.
\newblock To appear.

\bibitem{Gunduz:ISITA:08}
D.~Gunduz, R.~Brown, and {H. V. Poor}, ``Secret communication with feedback,''
  in {\em Proc. IEEE Intl. Symposium on Information Theory and its
  Applications}, (Auckland, New Zealand), Dec. 2008.

\bibitem{Khisti:TIT:07}
A.~Khisti and G.~W. Wornell, ``Secure transmission with multiple antennas: The
  {MISOME} wiretap channel,'' {\em IEEE Trans. Inform. Theory}, 2009.
\newblock To appear.

\bibitem{Oggier:TIT:07}
F.~Oggier and B.~Hassibi, ``The secrecy capacity of the {MIMO} wiretap
  channel,'' {\em IEEE Trans. Inform. Theory}, Oct. 2007.
\newblock Submitted.

\bibitem{Shafiee:TIT:09}
S.~Shafiee, N.~Liu, and S.~Ulukus, ``Towards the secrecy capacity of the
  {G}aussian {MIMO} wire-tap channel: The 2-2-1 channel,'' {\em IEEE Trans.
  Inform. Theory}, vol.~55, pp.~4033 -- 4039, Sept. 2009.

\bibitem{Liu:TIT:091}
R.~Liu and H.~V. Poor, ``Secrecy capacity region of a multi-antenna {G}aussian
  broadcast channel with confidential messages,'' {\em IEEE Trans. Inform.
  Theory}, vol.~55, pp.~1235--1249, Mar. 2009.

\bibitem{Lai:TIT:061}
L.~Lai and H.~{El Gamal}, ``The relay-eavesdropper channel: Cooperation for
  secrecy,'' {\em IEEE Trans. Inform. Theory}, vol.~54, pp.~4005--4019, Sept.
  2008.

\bibitem{Tekin:TIT:06}
E.~Tekin and A.~Yener, ``The {G}aussian multiple access wire-tap channel,''
  {\em IEEE Trans. Inform. Theory}, vol.~54, pp.~5747 -- 5755, Dec. 2008.

\bibitem{Liang:TIT:08}
Y.~Liang and H.~V. Poor, ``Multiple access channels with confidential
  messages,'' {\em IEEE Trans. Inform. Theory}, vol.~54, pp.~976--1002, Mar.
  2008.

\bibitem{Khisti:TIT:08}
A.~Khisti, A.~Tchamkerten, and G.~W. Wornell, ``Secure broadcasting over fading
  channels,'' {\em IEEE Trans. Inform. Theory}, vol.~54, pp.~2453--2469, June
  2008.

\bibitem{Bloch:TIT:08}
M.~Bloch, J.~Barros, M.~R.~D. Rodrigues, and S.~W. McLaughlin, ``Wireless
  information-theoretic security,'' {\em IEEE Trans. Inform. Theory}, vol.~54,
  pp.~2515--2534, June 2008.

\bibitem{Li:ITA:07}
Z.~Li, R.~Yates, and W.~Trappe, ``Secure communication over wireless
  channels,'' in {\em Proc. Inform. Theory and Appl. Workshop}, (La Jolla,
  CA.), Jan. 2007.

\bibitem{Simeone:CISS:09}
O.~Simeone and A.~Yener, ``The cognitive multiple access wire-tap channel,'' in
  {\em Proc. Conf. Inform. Science and Systems}, (Baltimore, MD), Mar. 2009.

\bibitem{Parada:ISIT:05}
P.~Parada and R.~Blahut, ``Secrecy capacity of {SIMO} and slow fading
  channels,'' in {\em Proc. IEEE Int. Symposium Inform. Theory}, (Adelaide,
  Australia), pp.~2152--2155, Sep. 2005.

\bibitem{Liu:ISIT:06}
R.~Liu, I.~Maric, R.~D. Yates, and P.~Spasojevic, ``The discrete memoryless
  multiple access channel with confidential messages,'' in {\em Proc. IEEE Int.
  Symposium Inform. Theory}, (Seattle, WA), pp.~957--961, July 2006.

\bibitem{Oohama:TIT:06}
Y.~Oohama, ``Relay channels with confidential messages,'' {\em IEEE Trans.
  Inform. Theory}, Nov. 2006.
\newblock Submitted.

\bibitem{Ekrem:TIT:09M}
E.~Ekrem and S.~Ulukus, ``The secrecy capacity region of the {G}aussian {MIMO}
  multi-receiver wiretap channel,'' {\em IEEE Trans. Inform. Theory}, Mar.
  2009.
\newblock Submitted.

\bibitem{Yuksel:ITW:09}
M.~Yuksel, X.~Liu, and E.~Erkip, ``A secure communication game with a relay
  helping the eavesdropper,'' in {\em Proc. IEEE Information Theory Workshop},
  (Taormina, Italy), Oct. 2009.
\newblock To appear.

\bibitem{Ali:TIFS:11}
S.~Ali, A.~Fakoorian, and A.~L. Swindlehurst, ``Mimo interference channel with
  confidential messages: Achievable secrecy rates and precoder design,'' {\em
  IEEE Trans. Inf. Forensics and Security}, vol.~6, pp.~640--649, Sep. 2011.

\bibitem{Li:TSP:11}
J.~Li, A.~P. Petropulu, and S.~Weber, ``On cooperative relaying schemes for
  wireless physical layer security,'' {\em IEEE Trans. Signal Processing},
  vol.~59, pp.~4985--4997, Oct. 2011.

\bibitem{Chen:TIFS:12}
J.~Chen, R.~Zhang, L.~Song, Z.~Han, and B.~Jiao, ``Joint relay and jammer
  selection for secure two-way relay networks,'' {\em IEEE Trans. Inf.
  Forensics and Security}, vol.~7, pp.~310--320, Feb. 2012.

\bibitem{Liu:ITW:07}
R.~Liu, Y.~Liang, H.~V. Poor, and P.~Spasojevic, ``Secure nested codes for type
  {II} wiretap channels,'' in {\em Proc. IEEE Information Theory Workshop},
  (Lake Tahoe, CA), Sep. 2-6 2007.

\bibitem{Bloch:ISIT:06}
M.~Bloch, A.~Thangaraj, S.~W. McLaughlin, and J.-M. Merolla, ``{LDPC} based
  secret key agreement over the gaussian wiretap channel,'' in {\em Proc. IEEE
  Int. Symposium Inform. Theory}, (Seattle, WA), pp.~1179 -- 1183, July 2006.

\bibitem{tassiulas}
L.~Tassiulas and A.~Ephremides, ``Jointly optimal routing and scheduling in
  packet ratio networks,'' {\em IEEE Transactions on Information Theory},
  vol.~38, pp.~165 --168, Jan. 1992.

\bibitem{shroff}
X.~Liu, E.~K.~P. Chong, and N.~B. Shroff, ``A framework for opportunistic
  scheduling in wireless networks,'' {\em Computer Networks}, vol.~41, no.~4,
  pp.~451--474, 2003.

\bibitem{subramanian}
J.~Huang, V.~Subramanian, R.~Agrawal, and R.~Berry, ``Downlink scheduling and
  resource allocation for ofdm systems,'' {\em IEEE Transactions on Wireless
  Communications}, vol.~8, no.~1, pp.~288 --296, 2009.

\bibitem{urgaonkar}
R.~Urgaonkar and M.~J. Neely, ``Opportunistic scheduling with reliability
  guarantees in cognitive radio networks,'' {\em IEEE Trans. Mob. Comput.},
  vol.~8, no.~6, pp.~766--777, 2009.

\bibitem{jaramillo}
J.~J. Jaramillo and R.~Srikant, ``Optimal scheduling for fair resource
  allocation in ad hoc networks with elastic and inelastic traffic,'' in {\em
  INFOCOM}, pp.~2231--2239, 2010.

\bibitem{stolyar}
A.~Stolyar, ``Greedy primal-dual algorithm for dynamic resource allocation in
  complex networks,'' {\em Queueing Systems}, vol.~54, pp.~203--220, 2006.
\newblock 10.1007/s11134-006-0067-2.

\bibitem{kelly}
F.~P. Kelly, A.~K. Maulloo, and D.~K.~H. Tan, ``{Rate Control for Communication
  Networks: Shadow Prices, Proportional Fairness and Stability},'' {\em The
  Journal of the Operational Research Society}, vol.~49, no.~3, pp.~237--252,
  1998.

\bibitem{Low}
S.~H. Low and D.~E. Lapsley, ``Optimization flow control-i: basic algorithm and
  convergence,'' {\em IEEE/ACM Trans. Netw.}, vol.~7, no.~6, pp.~861--874,
  1999.

\bibitem{Kar}
X.~Wang and K.~Kar, ``Cross-layer rate control for end-to-end proportional
  fairness in wireless networks with random access,'' in {\em MobiHoc},
  pp.~157--168, 2005.

\bibitem{Neely}
L.~Georgiadis, M.~J. Neely, and L.~Tassiulas, ``Resource allocation and
  cross-layer control in wireless networks,'' {\em Foundations and Trends in
  Networking}, vol.~1, no.~1, 2006.

\bibitem{Knopp:ICC:95}
R.~Knopp and P.~A. Humblet, ``Information capacity and power control in
  single-cell multiusercommunications,'' in {\em Proc. IEEE Int. Conf.
  Commun.}, vol.~1, (Seattle, WA), pp.~331--335, Jun. 18-22, 1995.

\bibitem{Wyner}
A.~D. Wyner, ``The wire-tap channel,'' {\em The Bell System Technical Journal},
  vol.~54, pp.~1355--1387, Oct. 1975.

\bibitem{bc-secrecy}
I.~Csisz$\acute{\text{a}}$r and J.~K$\ddot{\text{o}}$rner, ``Broadcast channels
  with confidential messages,'' {\em IEEE Trans. Inform. Theory}, vol.~24,
  pp.~339--348, May 1978.

\bibitem{tech_report_Koksal_Ercetin}
C.~E. Koksal and O.~Ercetin, ``Control of wireless networks with secrecy,''
  technical report, 2010.
\newblock http://arxiv.org/abs/cs/1101.3444.

\bibitem{Tse:Book:05}
D.~Tse and P.~Viswanath, {\em Fundamentals of Wireless Communication}.
\newblock New York: Cambridge University Press, 2005.

\bibitem{Frenger}
P.~Frenger, ``Turbo decoding for wireless systems with imperfect channel
  estimates,'' {\em IEEE Transactions on Communications}, vol.~48, no.~9,
  pp.~1437 --1440, 2000.

\bibitem{Kok_Sch_10}
C.~E. Koksal and P.~Schniter, ``Robust rate-adaptive wireless communication
  using ack/nak-feedback,'' {\em IEEE Trans. on Signal Processing}, 2012.
\newblock to appear.

\bibitem{Neely:TIT}
M.~J. Neely, ``Energy optimal control for time-varying wireless networks,''
  {\em IEEE Transactions on Information Theory}, vol.~52, no.~7,
  pp.~2915--2934, 2006.

\end{thebibliography}

\appendices

{\allowdisplaybreaks
\vspace{-0.2in}

\section{Proof of Theorem~\ref{the:pri_opp_sched_ach_up}}
\label{sec:appen_proof_0}

Let us further introduce the following notation:

\noindent $W_j^\text{rand}$: randomization sequence associated with message $W_j^\text{priv}$,

\noindent $\vc{X}(k)$: transmitted vector of ($N_1$) symbols over block $k$,

\noindent $\vc{X}_j=\{\vc{X}(k)|{\cal I}^{\text{POS}}_j(k)=1 \}$: the transmitted signal over block $k$, whenever ${\cal I}^{\text{POS}}_j(k)=1$ (i.e., node $j$ is the active transmitter)

\noindent $\vc{Y}_i(k)$: the received vector of symbols at node $i$ ($\vc{Y}_b(k)$ for the base station) over block $k$,

\noindent $\vc{Y}_i^j=\{\vc{Y}_i(k)|{\cal I}^{\text{POS}}_j(k)=1 \}$: the received
signal at node $i$ over block $k$, whenever ${\cal I}^{\text{POS}}_j(k)=1$ (i.e., node
$j$ is the active transmitter). We use $\vc{Y}_j^j$ for the received signal by the base station.

The equivocation analysis follows directly for the described privacy
scheme: For any given node $j$, we have
\begin{align}
\label{eq:ach_0}
H(W_j^\text{priv}|\vc{Y}_i^j) &\geq H(W_j^\text{priv}|\vc{Y}_{i^*(j)}^j) \\
\nonumber &= I(W_j^\text{priv};\vc{Y}^j_1,\ldots,\vc{Y}^j_n|\vc{Y}^j_{i^*(j)}) +
H(W_j^\text{priv}|\vc{Y}^j_1,\ldots,\vc{Y}^j_n) \\
\nonumber
&\geq I(W_j^\text{priv};\vc{Y}^j_1,\ldots,\vc{Y}^j_n|\vc{Y}^j_{i^*(j)}) \\
\nonumber &=
I(W_j^\text{priv},W_j^\text{rand};\vc{Y}^j_1,\ldots,\vc{Y}^j_n|\vc{Y}^j_{i^*(j)}) \\
\label{eq:ach_1}
&\hspace{0.3in} - I(W_j^\text{rand};\vc{Y}^j_1,\ldots,\vc{Y}^j_n|\vc{Y}^j_{i^*(j)},W_j^\text{priv}) \\
\nonumber &=
I(W_j^\text{priv},W_j^\text{rand};\vc{Y}^j_1,\ldots,\vc{Y}^j_n|\vc{Y}^j_{i^*(j)})\\
\nonumber &\hspace{0.3in}-
H(W_j^\text{rand}|\vc{Y}^j_{i^*(j)},W_j^\text{priv}) \\
\nonumber &\hspace{0.3in} + H(W_j^\text{rand}|\vc{Y}^j_1,\ldots,\vc{Y}^j_n,W_j^\text{priv}) \\
\nonumber &\geq
I(W_j^\text{priv},W_j^\text{rand};\vc{Y}^j_1,\ldots,\vc{Y}^j_n|\vc{Y}^j_{i^*(j)})\\
\nonumber&\hspace{0.3in}-
H(W_j^\text{rand}|\vc{Y}^j_{i^*(j)},W_j^\text{priv}) \\
\label{eq:ach_3} &\geq
I(W_j^\text{priv},W_j^\text{rand};\vc{Y}^j_1,\ldots,\vc{Y}^j_n|\vc{Y}^j_{i^*(j)}) -
N\varepsilon_1 \\
\nonumber
&=I(\vc{X}_j;\vc{Y}^j_1,\ldots,\vc{Y}^j_n|\vc{Y}^j_{i^*(j)}) \\
\label{eq:ach_4}
&\hspace{0.2in} -
I(\vc{X}_j;\vc{Y}^j_1,\ldots,\vc{Y}^j_n|\vc{Y}^j_{i^*(j)},W_j^\text{priv},W_j^\text{rand}) - N\varepsilon_1 \\
\label{eq:ach_5} &\geq
I(\vc{X}_j;\vc{Y}^j_1,\ldots,\vc{Y}^j_n|\vc{Y}^j_{i^*(j)}) -
N(\varepsilon_1+\varepsilon_2) \\
\label{eq:ach_6} &= I(\vc{X}_j;\vc{Y}^j_1,\ldots,\vc{Y}^j_n) -
I(\vc{X}_j;\vc{Y}^j_{i^*(j)}) -
N(\varepsilon_1+\varepsilon_2) \\
\label{eq:ach_6.5} &\geq I(\vc{X}_j;\vc{Y}^j_j) -
I(\vc{X}_j;\vc{Y}^j_{i^*(j)}) - N(\varepsilon_1+\varepsilon_2) \\
\nonumber &= \sum_{k:{\cal I}^{\text{POS}}_j(k)=1} \hspace{-0.1in}\left[ I(\vc{X}(k);\vc{Y}_j(k)) - I(\vc{X}(k);\vc{Y}_{i^*(j)}(k)) \right] \\
\label{eq:ach_7}
&\hspace{1in} - N(\varepsilon_1+\varepsilon_2) \\
\label{eq:ach_8} &\geq N\left[p^M_j \left((\bar{R}_j^M - \delta) - \bar{R}_j^m
\right) - (\varepsilon_1+\varepsilon_2+\varepsilon_3)\right]
\end{align}
with probability 1, for any positive
$(\varepsilon_1,\varepsilon_2,\varepsilon_3)$ triplet and
arbitrarily small $\delta$, as $N_1,N_2$ go to $\infty$. Here,
(\ref{eq:ach_0}) follows since $i^*(j)=\argmax_{i \in \{1,\ldots ,n\}} I(W_j^\text{priv};\vc{Y}_i^j)$ \textr{($W_j^\text{priv} \leftrightarrow \vc{X}_j \leftrightarrow \vc{Y}_{i^*(j)}^j \leftrightarrow \vc{Y}_i^j$ forms a Markov chain for all $i$ and data processing inequality)}, (\ref{eq:ach_1}) is by the chain rule,
(\ref{eq:ach_3}) follows from the application of Fano's inequality
(as we choose the rate of the randomization sequence to be
$N(\bar{R}_j^m-\delta) < I(W_j^\text{rand};\vc{Y}^j_{i^*(j)})$, which allows for the
randomization message to be decoded at node $i^*(j)$, given the bin
index), (\ref{eq:ach_4}) follows from the chain rule and that
$(W_j^\text{priv},W_j^\text{rand}) \leftrightarrow \vc{X}_j \leftrightarrow
(\vc{Y}_1^j,\ldots ,\vc{Y}_n^j)$ forms a Markov chain,
(\ref{eq:ach_5}) holds since
$I(\vc{X}_j;\vc{Y}_1^j,\ldots,\vc{Y}_n^j|\vc{Y}_{i^*(j)}^j,W_j^\text{priv},W_j^\text{rand})
\leq N\varepsilon_2$ as the transmitted symbol sequence
$\vc{X}_j$ is determined w.p.1 given $(\vc{Y}_{i^*(j)}^j,W_j^\text{priv},W_j^\text{rand})$,
(\ref{eq:ach_6}) follows from the chain rule, (\ref{eq:ach_6.5}) holds since $\vc{Y}_j^j(k)$ is an entry of vector $[\vc{Y}_1^j(k),\ldots ,\vc{Y}_n^j(k)]$, (\ref{eq:ach_7}) holds because the fading processes are iid, and finally
(\ref{eq:ach_8}) follows from strong law of large numbers.

Thus, with the described privacy scheme, the perfect privacy constraint is satisfied for all nodes, since for any $j \in \{1,\ldots ,n\}$, we have
\begin{align}
\nonumber
\frac{1}{N}I(W_j^\text{priv};\vc{Y}_i^j) &= \frac{1}{N} (H(W_j^\text{priv})-H(W_j^\text{priv}|\vc{Y}_i^j)) \\
\nonumber
&\leq R_j^{\text{priv}}-[p^M_j\left( (\bar{R}_j^M-\delta)
- \bar{R}_j^m \right) - (\varepsilon_1+\varepsilon_2+\varepsilon_3)] \\
\label{eq:equiv}
&\leq \varepsilon,
\end{align}
for any given $\varepsilon>0$. We just showed that, with private opportunistic scheduling, a private information rate of $R_j^{\text{priv}}=p^M_j(\bar{R}_j^M-\bar{R}_j^m)$ is achievable for any given node $j$.
\endproof

\vspace{-0.15in}
\section{Proof of Theorem~\ref{the:pri_opp_sched_uplink}}
\label{sec:appen_proof_1}

 The proof uses the notation introduced in the first paragraph of Appendix~\ref{sec:appen_proof_0}. To meet the perfect secrecy constraint, it is necessary and sufficient to guarantee $\lim_{N\rightarrow \infty} \frac{1}{N} I(W_j^\text{priv};\vc{Y}_{i^*(j)}^j) \leq \varepsilon$ for all nodes $j\in \{1,\ldots ,n\}$. Since $n<\infty$, one can write an equivalent condition on the sum mutual information over each node:
\begin{align}
\nonumber
\varepsilon' &\geq \frac{1}{N} \sum_{j=1}^{n} I(W_j^\text{priv};\vc{Y}_{i^*(j)}^j) \\
\nonumber &= \frac{1}{N} \sum_{j=1}^{n} \left[ H(W_j^\text{priv})- H(W_j^\text{priv}|\vc{Y}_{i^*(j)}^j) \right] \\
\label{eq:conv_0}
&= R_{\text{sum}}^{\text{priv}} - \frac{1}{N} \sum_{j=1}^{n} H(W_j^\text{priv}|\vc{Y}_{i^*(j)}^j) \\
\nonumber &= R_{\text{sum}}^{\text{priv}} - \frac{1}{N}
\sum_{j=1}^{n} \left[
I(W_j^\text{priv};\vc{Y}_1^j,\ldots,\vc{Y}_n^j|\vc{Y}_{i^*(j)}^j)
\right.\\
\nonumber&\hspace{1.25in}\left.+
H(W_j^\text{priv}|\vc{Y}_1^j,\ldots,\vc{Y}_n^j) \right] \\
\nonumber &\geq R_{\text{sum}}^{\text{priv}} - \frac{1}{N} \sum_{j=1}^{n} \left[
I(W_j^\text{priv},W_j^\text{rand};\vc{Y}_1^j,\ldots,\vc{Y}_n^j|\vc{Y}_{i^*(j)}^j) \right. \\
\label{eq:conv_1}
&\hspace{0.6in} -
\left. I(W_j^\text{rand};\vc{Y}_1^j,\ldots,\vc{Y}_n^j|\vc{Y}_{i^*(j)}^j,W_j^\text{priv}) + N\varepsilon_4 \right] \\
\nonumber &\geq R_{\text{sum}}^{\text{priv}} - \frac{1}{N} \sum_{j=1}^{n} \left[
I(W_j^\text{priv},W_j^\text{rand};\vc{Y}_1^j,\ldots,\vc{Y}_n^j|\vc{Y}_{i^*(j)}^j) +
N\varepsilon_4 \right] \\
\label{eq:conv_2} &\geq R_{\text{sum}}^{\text{priv}} - \frac{1}{N} \sum_{j=1}^{n} \left[
I(\vc{X}_j;\vc{Y}_1^j,\ldots,\vc{Y}_n^j|\vc{Y}_{i^*(j)}^j) +
N\varepsilon_4 \right] \\
\label{eq:conv_4} &= R_{\text{sum}}^{\text{priv}} - \frac{1}{N} \sum_{j=1}^{n} \left[ I(\vc{X}_j;\vc{Y}_1^j,\ldots,\vc{Y}_n^j) -
I(\vc{X}_j;\vc{Y}_{i^*(j)}^j) + N\varepsilon_4 \right] \\
\nonumber &= R_{\text{sum}}^{\text{priv}} - \frac{1}{N} \sum_{j=1}^{n} \left[ I(\vc{X}_j;\vc{Y}_j^j) + I(\vc{X}_j;\vc{Y}_1^j,\ldots,\vc{Y}_n^j|\vc{Y}_j^j) \right. \\
\label{eq:conv_4.5_1}
&\hspace{0.6in} -
\left. I(\vc{X}_j;\vc{Y}_{i^*(j)}^j) + N\varepsilon_4 \right] \\
\nonumber &\hspace{-0.1in} \geq R_{\text{sum}}^{\text{priv}} - \frac{1}{N} \sum_{j=1}^{n} \left[ I(\vc{X}_j;\vc{Y}_j^j) + H(\vc{X}_j|\vc{Y}_j^j) - I(\vc{X}_j;\vc{Y}_{i^*(j)}^j) + N\varepsilon_4 \right] \\
\label{eq:conv_4.5_2}
&\hspace{-0.1in} \geq R_{\text{sum}}^{\text{priv}} - \frac{1}{N} \sum_{j=1}^{n} \left[ I(\vc{X}_j;\vc{Y}_j^j) + H(W_j^\text{priv}|\vc{Y}_j^j) - I(\vc{X}_j;\vc{Y}_{i^*(j)}^j) + N\varepsilon_4 \right] \\
\label{eq:conv_4.5_3}
&\geq R_{\text{sum}}^{\text{priv}} - \frac{1}{N} \sum_{j=1}^{n} \left[ I(\vc{X}_j;\vc{Y}_j^j) - I(\vc{X}_j;\vc{Y}_{i^*(j)}^j) + N(\varepsilon_4+\varepsilon_5) \right] \\
\nonumber &= R_{\text{sum}}^{\text{priv}} - \frac{1}{N} \sum_{j=1}^{n}\left\{ \sum_{k:{\cal I}_j(k)=1} \left[ I(\vc{X}(k) ; \vc{Y}_b(k)) \right. \right. \\
\label{eq:conv_5}
&\hspace{1.1in}  \left. \left. - I(\vc{X}(k) ; \vc{Y}_{i^*(j)}(k)) \right] + \varepsilon_4 + \varepsilon_5 \right\} \\
\nonumber &\geq R_{\text{sum}}^{\text{priv}} -  \frac{1}{N} \sum_{j=1}^{n} \max_{{\cal I}_j(k)}\left\{ \sum_{k:{\cal I}_j(k)=1} \left[ I(\vc{X}(k) ; \vc{Y}_b(k)) \right. \right. \\
\nonumber
&\hspace{1.25in} \left. \left. - I(\vc{X}(k) ; \vc{Y}_{i^*(j)}(k)) \right] +  \varepsilon_4 + \varepsilon_5 \right\} \\
\nonumber &= R_{\text{sum}}^{\text{priv}} -  \frac{1}{N} \sum_{j=1}^{n}\left\{ \sum_{k:{\cal I}_j^{\text{POS}}(k)=1} \left[ I(\vc{X}(k) ; \vc{Y}_b(k)) \right. \right. \\
\label{eq:conv_5.5_1}
&\hspace{1.1in} \left. \left. - I(\vc{X}(k) ; \vc{Y}_{i^*(j)}(k)) \right] +  \varepsilon_4 + \varepsilon_5 \right\} \\
\label{eq:conv_6} &\geq R_{\text{sum}}^{\text{priv}} - \sum_{j=1}^{n} \left[p^M_j \left(\bar{R}_j^M - \bar{R}_j^m \right) + \varepsilon_4+\varepsilon_5+\varepsilon_6 \right]
\end{align}
with probability 1, for any positive $\varepsilon'$ and $(\varepsilon_4,\varepsilon_5,\varepsilon_6)$
triplet as $N_1,N_2$ go to $\infty$. Here, (\ref{eq:conv_0}) follows from the definition of $R_{\text{sum}}^{\text{priv}}$ and that $\frac{1}{N}H(W_j^\text{priv})=R_j^{\text{priv}}$; (\ref{eq:conv_1}) follows from the chain rule and Fano's inequality (as $H(W_j^\text{priv}|\vc{Y}_1^j,\ldots,\vc{Y}_n^j) \leq
H(W_j^\text{priv},W_j^\text{rand}|\vc{Y}_1^j,\ldots,\vc{Y}_n^j) \leq N\varepsilon_4$
since the message pair $(W_j^\text{priv},W_j^\text{rand})$ can be decoded with
arbitrarily low probability of error given $(\vc{Y}_1^j,\ldots
,\vc{Y}_n^j)$); (\ref{eq:conv_2}) is from the data processing
inequality as $(W_j^\text{priv},W_j^\text{rand})\leftrightarrow \vc{X}_j \leftrightarrow
(\vc{Y}_1^j,\ldots ,\vc{Y}_n^j)$ forms a Markov chain;
(\ref{eq:conv_4}) and (\ref{eq:conv_4.5_1}) follow from the chain rule; (\ref{eq:conv_4.5_2}) follows from the data processing inequality; (\ref{eq:conv_4.5_3}) follows since node $j$ decodes message $W_j^\text{priv}$ with arbitrarily low probability of error $\varepsilon_5$; (\ref{eq:conv_5}) holds since the fading processes are iid; (\ref{eq:conv_5.5_1}) holds because private opportunistic scheduler chooses ${\cal I}_j^{\text{POS}}(k)=\argmax_{{\cal I}_j(k)} [ R_j(k)-R_{ji^*(j)}(k) ] = \argmax_{{\cal I}_j(k)} \left[ I(\vc{X}(k) ; \vc{Y}_b(k)) - I(\vc{X}(k) ; \vc{Y}_{i^*(j)}(k)) \right]$ for all $k$; and finally (\ref{eq:conv_6}) follows by an application of the strong law of large numbers. The above derivation leads to the desired result:
\begin{equation}
\label{eq:ach_sum_rate}
R_{\text{sum}}^{\text{priv}} \leq \sum_{j=1}^n \left[ p^M_j \left(\bar{R}_j^M - \bar{R}_j^m \right) \right] .
\end{equation}
We complete the proof noting that the above sum rate is achievable by private opportunistic scheduling as shown in (\ref{eq:ach_8}).
\endproof

Note that, from the above steps, we can also see that the individual private information rates given in Theorem~\ref{the:pri_opp_sched_ach_up} are the maximum achievable individual rates with private opportunistic scheduling. This is due to the fact that, for any node $j$, with private opportunistic scheduling, the above derivation lead to:
\begin{equation}
\label{eq:up_bnd}
\frac{1}{N} H(W_j^\text{priv}|\vc{Y}_{i^*(j)}^j) \leq p^M_j \left(\bar{R}_j^M - \bar{R}_j^m \right) + \varepsilon
\end{equation}
for any $\varepsilon > 0$ as $N\rightarrow \infty$. Consequently, with private opportunistic scheduling, no node can achieve any individual privacy rate above that given in (\ref{eq:up_bnd}), hence the converse of Theorem~\ref{the:pri_opp_sched_ach_up} also holds.

\vspace{-0.2in}\section{Proof of Lemma \ref{lemma:drift-1}}
\label{proof:drift-1} \proof Since the maximum transmission power is
finite, in any interference-limited system transmission rates are
bounded. Let $R^{p,max}_j$ and $R^{o,max}_j$ be the maximum private
and open rates for user $j$, which depends on the channel states.
Also assume that the arrival rates are bounded, i.e., $A^{p,max}_j$
and $A^{o,max}_i$ be the maximum number of private and open bits
that may arrive in a block for each user. Hence, the following
inequalities can be obtained for each private queue:\textr{
\begin{align}
(Q^p_j&(k+1))^2-(Q^p_j(k))^2 \nonumber\\
&=\left(\left[
Q_j^p(k)-R_j^p(k)\right]^+ +A_j^p(k)\right)^2-(Q_j^p(k))^2 \nonumber\\
&\leq(Q^p_j(k))^2+(A^p_j(k))^2+(R^p_j(k))^2\nonumber\\
&\hspace{0.4cm}-2Q_j^p(k)\left[R_j^p(k)-A_j^p(k)\right]-(Q^p_j(k))^2 \nonumber \\
&\leq (R^p_j(k))^2 + (A^p_j(k))^2-2Q^p_j(k)[R^p_j(k)-A^p_j(k)]\nonumber \\
&\leq B_1-2Q^p_j(k)[R^p_j(k)-A^p_j(k)]\label{eq:Qs}
\end{align}
where  $B_1 =(R^{p,max}_j)^2+(A^{p,max}_j)^2$. The same line of
derivation can be performed for open queues to obtain:
\begin{align}
(Q^o_j&(k+1))^2-(Q^o_j(k))^2\nonumber\\
&= \left(\left[ Q_j^o(k)-R_j^o(k)\right]^+ + A_j^o(k)\right)^2
-(Q_j^o(k))^2 \nonumber\\
&\leq B_2-2Q^o_j(k)[R^o_j(k)-A^o_j(k)] \label{eq:Qm}
\end{align}
where $B_2 = (R^{o,max}_j)^2+(A^{o,max}_j)^2$ }.

Hence, by taking expectation, multiplying by $\frac{1}{2}$, and
summing \eqref{eq:Qs}-\eqref{eq:Qm} over all $j=1,\ldots,n$, we
obtain the upper bound on $\Delta(k)$ as given in the Lemma, where
$B=n(B_1+B_2)/2$.
\endproof

\vspace{-0.15in}
\section{Proof of Theorem \ref{thm:optimalcontrol-1}}
\label{proof:optimalcontrol-1}

\proof Lyapunov Optimization Theorem \cite{Neely} suggests that a
good control strategy is the one that minimizes the following:
\begin{equation}
\Delta^U(k)=\Delta(k) - V \E{\sum_j
\left(g^p_j(k)+g^o_j(k)\right)\Bigm\vert
\mathbf{Q^p(k)},\mathbf{Q^o(k)}} \label{eq:deltawithreward}
\end{equation}

By using \eqref{eq:delta}, we may obtain an upper bound for
\eqref{eq:deltawithreward}, as follows:
\begin{align}
\Delta^U(k)< B&-\sum_j \mathbb{E}\left[
Q^p_j(k)[R^p_j(k)-A^p_j(k)]\bigm\vert Q^p_j(k)\right]\nonumber\\
&-\sum_j \mathbb{E}\left[ Q^o_j(k)[R^o_j(k)-A^o_j(k)]\bigm\vert Q^o_j(k)\right] \nonumber\\
& -V \E{\sum_j U^p_j(A_j^p(k))+\sum_j U^o_j(A_j^o(k))}
\label{drift_final}
\end{align}

By rearranging the terms in \eqref{drift_final} it is easy to
observe that our proposed dynamic network control algorithm
minimizes the right hand side of \eqref{drift_final}.

If the private and open arrival rates are in the feasible region,
it has been shown in \cite{Neely:TIT} that there must exist a
stationary scheduling and rate control policy that chooses the users
and their transmission rates independent of queue backlogs and only
with respect to the channel statistics.  In particular, the optimal
stationary policy can be found as the solution of a deterministic
policy if  a priori channel statistics are known.

Let $U^*$ be the optimal value of the objective function of the
problem
\eqref{eq:opt-objective-1}-\eqref{eq:const-stability-full-csi-2}
obtained by the aforementioned stationary policy. Also let
${\lambda_j^p}^*$ and ${\lambda_j^o}^*$ be optimal private and open
traffic arrival rates found as the solution of the same problem. In
particular, the optimal input rates ${\lambda_j^p}^*$ and
${\lambda_j^o}^*$ could in principle be achieved by the simple
backlog-independent admission control algorithm of including all new
arrivals $(A_j^p(k),A_j^o(k))$ for a given node $j$ in block $k$
independently with probability
$(\zeta_j^p,\zeta^o_j)=({\lambda_j^p}^*/\lambda_j^p,{\lambda_j^o}^*/\lambda_j^o)$.
\textr{ Then, the right hand side (RHS) of \eqref{drift_final} can
be rewritten as
\begin{align}
B&-\sum_j \mathbb{E}\left[
Q^p_j(k)\right]\mathbb{E}\left[R^p_j(k)-A^p_j(k)\right]\nonumber\\
&-\sum_j \mathbb{E}\left[
Q^o_j(k)\right]\mathbb{E}\left[R^o_j(k)-A^o_j(k)\right]  -V
U^*.\label{eq:rhs-drift-indep}
\end{align}}
Also, since $({\lambda_j^p}^*,{\lambda_j^o}^*)\in \Lambda$, i.e.,
arrival rates are strictly interior of the rate region, there must
exist a stationary scheduling and rate allocation policy that is
independent of queue backlogs and satisfies the following:
\begin{align}
\E{R_j^p\bigm\vert \mathbf{Q^p}} &\geq
{\lambda_j^p}^*+\epsilon_1 \label{eq:opt-conds-1}\\
\E{R_j^o\bigm\vert \mathbf{Q^o}} &\geq {\lambda_j^o}^*+\epsilon_2
\label{eq:opt-conds-2}
\end{align}


Clearly, any stationary policy should satisfy \eqref{drift_final}.
Recall that our proposed policy minimizes RHS of
\eqref{drift_final}, and hence, any other stationary policy
(including the optimal policy) has a higher RHS value than the one
attained by our policy. In particular, the stationary policy that
satisfies \eqref{eq:opt-conds-1}-\eqref{eq:opt-conds-2}, and
implements aforementioned probabilistic admission control can be
used to obtain an upper bound for the RHS of our proposed policy.
Inserting \eqref{eq:opt-conds-1}-\eqref{eq:opt-conds-2} into
\eqref{eq:rhs-drift-indep}, we obtain the following upper bound for
our policy:
\begin{align}
RHS<B&-\sum_j\epsilon_1\mathbb{E}[Q_j^p(k)]-\sum_j\epsilon_2\mathbb{E}[Q_j^o(k)]-VU^*.
\end{align}
This is exactly in the form of Lyapunov Optimization Theorem given
in Theorem \ref{thm:lyap}, and hence, we can obtain bounds on the
performance of the proposed policy and the sizes of queue backlogs
as given in Theorem \ref{thm:optimalcontrol-1}.
\endproof

%

}
\vspace{-0.5in}

\begin{IEEEbiography}[{\includegraphics[width=1in,height=1.25in,clip,keepaspectratio]{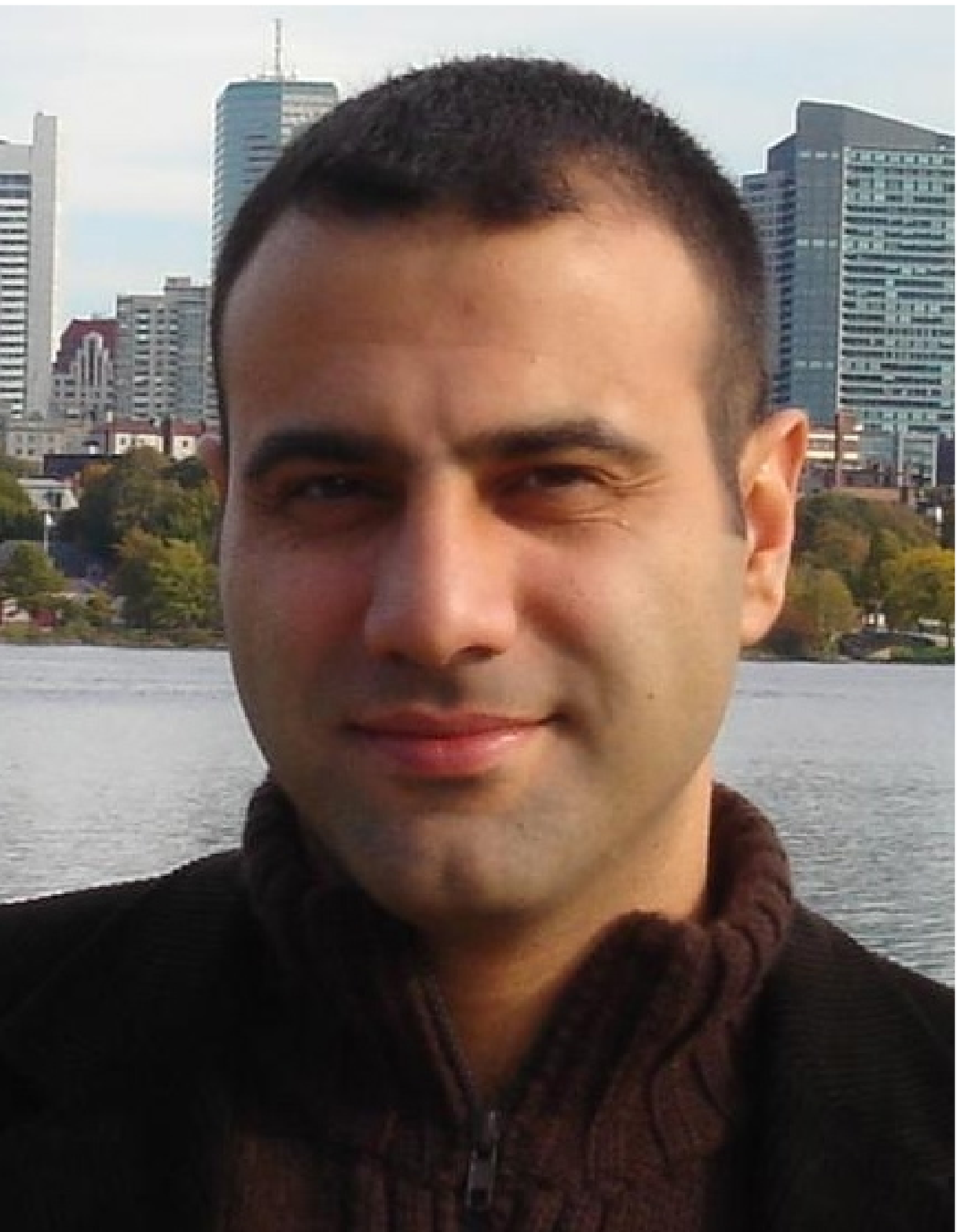}}]
{C.~Emre Koksal}
C. Emre Koksal received the B.S. degree in electrical engineering from the Middle East Technical
University, Ankara, Turkey, in 1996, and the S.M. and Ph.D. degrees from the Massachusetts Institute
of Technology (MIT), Cambridge, in 1998 and 2002, respectively, in electrical engineering and computer science. He was a Postdoctoral Fellow in the Networks and Mobile Systems Group in the Computer Science and Artificial Intelligence Laboratory, MIT and a Senior Researcher jointly in the Laboratory for Computer Communications and the Laboratory for Information Theory at EPFL, Lausanne, Switzerland. Since 2006, he has been an Assistant Professor in the Electrical and Computer Engineering Department, Ohio State University, Columbus, Ohio. His general areas of interest are wireless communication, communication networks, information theory, stochastic processes, and financial economics.

He is the recipient of the National Science Foundation CAREER Award (2011), the OSU College of Engineering Lumley Research Award (2011), and the co-recipient of an HP Labs - Innovation Research Award. The paper he co-authored was a best student paper candidate in MOBICOM 2005.
\end{IEEEbiography}

\vspace{-0.5in}
\begin{IEEEbiography}[{\includegraphics[width=1in,height=1.25in,clip,keepaspectratio]{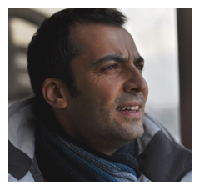}}]
{Ozgur Ercetin} received the BS degree in electrical and electronics
engineering from the Middle East Technical University, Ankara,
Turkey, in 1995 and the MS and PhD degrees in electrical engineering
from the University of Maryland, College Park, in 1998 and 2002,
respectively. Since 2002, he has been with the Faculty of Engineering and Natural
Sciences, Sabanci University, Istanbul. He was also a visiting
researcher at HRL Labs, Malibu, CA, Docomo USA Labs, CA, and The
Ohio State University, OH.  His research interests are in the field
of computer and communication networks with emphasis on fundamental
mathematical models, architectures and protocols of wireless
systems, and stochastic optimization.
\end{IEEEbiography}

\vspace{-0.5in}
\begin{IEEEbiography}[{\includegraphics[width=1in,height=1.25in,clip,keepaspectratio]{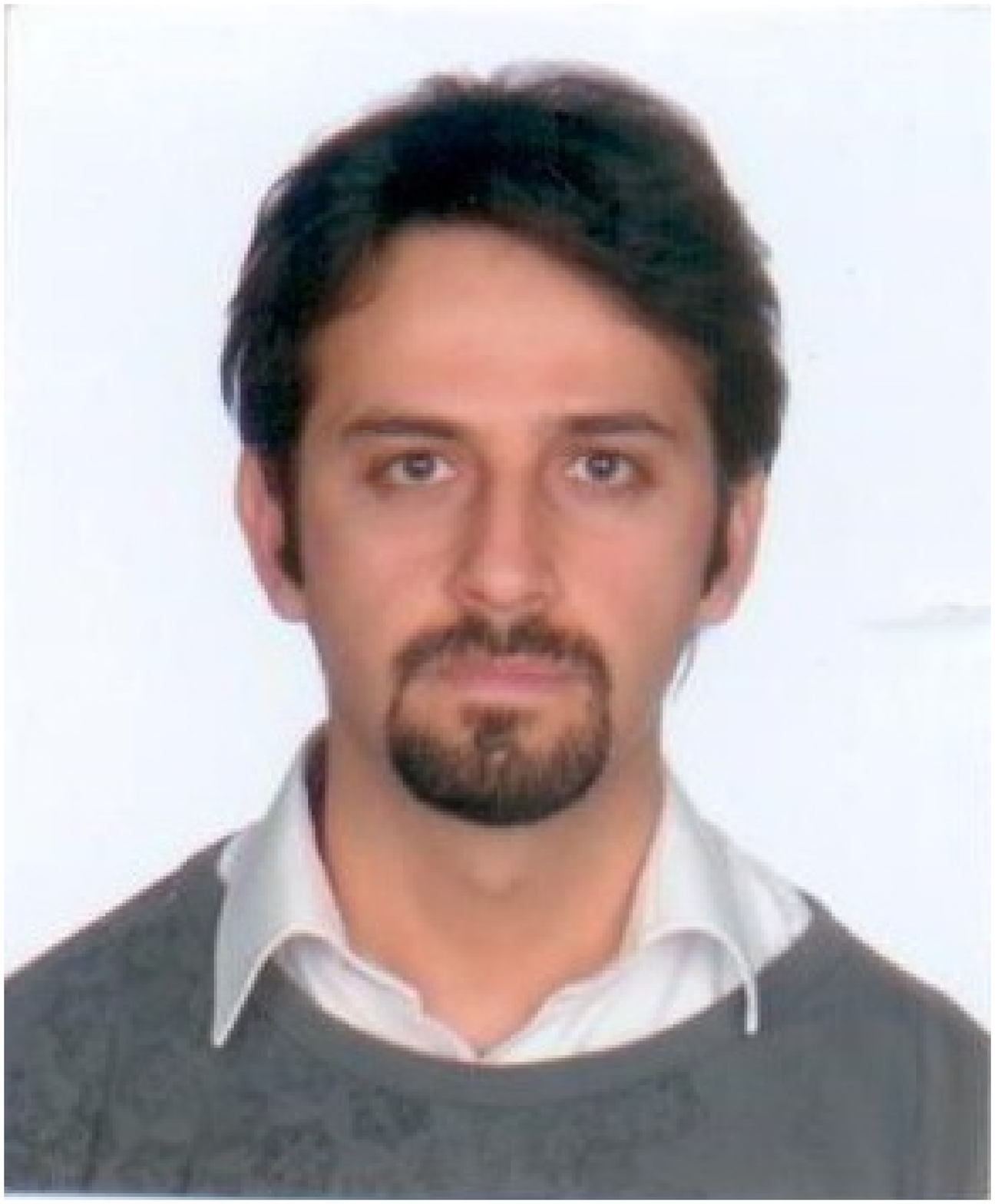}}]
{Yunus Sarikaya} received the BS and MS degrees in telecommunications
engineering from Sabanci University, Istanbul, Turkey, in 2006 and
2008, respectively. He is currently PhD student in electrical
engineering at Sabanci University.

His research interests include  optimal control of wireless
networks, stochastic optimization and information theoretical
security.
\end{IEEEbiography}

\end{document}